\documentclass[journal=jacsat,manuscript=article]{achemso}
\usepackage[version=3]{mhchem} % Formula subscripts using \ce{}
\usepackage{graphicx}
\author{Harish Gudla}
\affiliation{Department of Chemistry—Ångström Laboratory, Uppsala University, Lägerhyddsvägen 1, Box 538, 75121 Uppsala, Sweden}
\author{Kristina Edström}
\affiliation{Department of Chemistry—Ångström Laboratory, Uppsala University, Lägerhyddsvägen 1, Box 538, 75121 Uppsala, Sweden}
\author{Chao Zhang}
\affiliation{Department of Chemistry—Ångström Laboratory, Uppsala University, Lägerhyddsvägen 1, Box 538, 75121 Uppsala, Sweden}
\email{chao.zhang@kemi.uu.se}

\title[]
  {Salt effects on the mechanical properties of ionic conductive polymer: a molecular dynamics study}

\begin{document}

\begin{tocentry}
%\begin{figure}[ht]
  \includegraphics[width=1\textwidth]{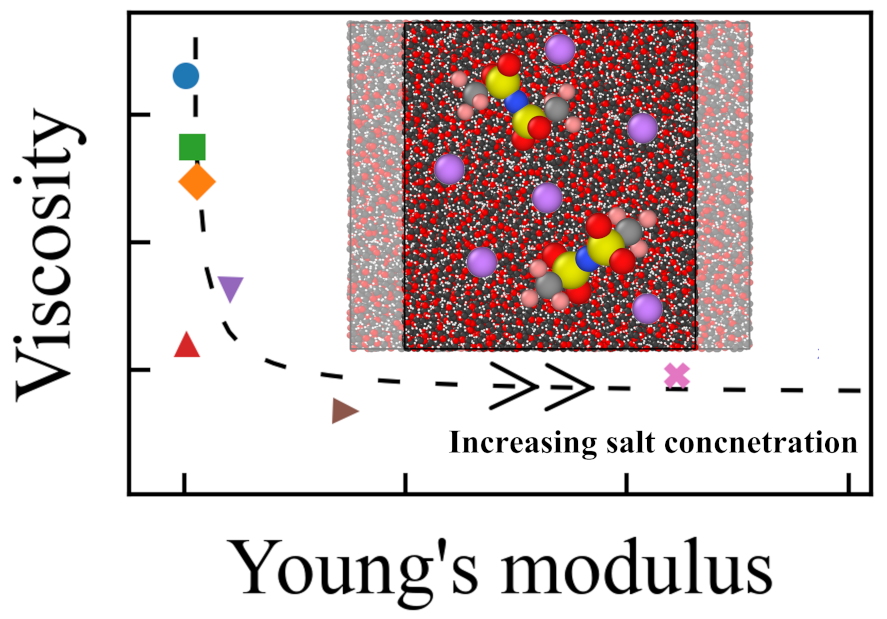}
%  \caption{}
% \label{fgr:toc}
%\end{figure}
\end{tocentry}

%Inside the \texttt{tocentry} environment, the font used is Helvetica
%8\,pt, as required by \emph{Journal of the American Chemical

%The surrounding frame is 9\,cm by 3.5\,cm, which is the maximum
%permitted for  \emph{Journal of the American Chemical Society}
%graphical table of content entries. The box will not resize if the
%content is too big: instead it will overflow the edge of the box.

\begin{abstract}
Functoinal polymers can be used as electrolyte and binder materials in solid-state batteries. This often requires performance targets in terms of both transport and mechanical properties.  In this work, a model ionic conductive polymer system, i.e., poly(ethylene oxide)-LiTFSI, was used to study the impact of salt concentrations on mechanical properties, including different types of elastic moduli and the visoelasticity with both non-equilibrium and equilibrium molecular dynamics simulations. We found an encouragingly good agreement between experiments and simulations regarding the Young's modulus, bulk modulus and viscosity. In addition, we identified an intermediate salt concentration at which the system shows high ionic conductivity, high Young's modulus and short elastic restoration time. Therefore, this study laid down the groundwork for investigating ionic conductive polymer binders with self-healing functionality from molecular dynamics simulations.
\end{abstract}

\section{Introduction}

Electrochemical energy storage, in particular, batteries, are key enablers for the green energy transition and the deployment of electric vehicles. This has led to an ever-increasing activities from both academia and industry with focuses on discovering new battery materials and cell chemistry leading to much higher energy density. However, due to the complexity of novel materials, they can be difficult to implemented in battery products at scale. To address this issue, large-scale research initiatives, e.g. the BATTERY 2030+~\cite{bat2030}, have identified thematic areas, e.g. integrating smart functionalities of sensing and self-healing~\cite{Narayan.2021}, into to the battery design. 

One of the promising approach to implement the self-healing functionality into the next generation anode materials (Si as a prominent example) is to explore functional polymers as binder materials~\cite{Liu.2017, Yang.2021dnr, Zhou22}. For example, functional groups that involve hydrogen bonding~\cite{chen19, Mi.2019, Xie2021} is a popular choice for polymer binders that can mitigate the large volume expansion of Si anodes during the cycling. Other types of self-healing mechanisms~\cite{yang15, Li22}, such dynamical covalent bonds~\cite{2019.Dufort}, ionic cross-linking~\cite{zhang14} and host-guest interactions~\cite{Kwon.2015}, are also interesting options. 

In all these cases, an understanding of the mechanical properties of ionic conductive polymer is necessary.  Besides its electrochemical stability, a good ionic conductive polymer should satisfy the requirement of both ionic conductivity ($\geq 10^{-5}$ S cm$^{-1}$ at 25 $^{\circ}$C) and mechanical strength ($\geq  30$ MPa at 25 $^{\circ}$C)~\cite{Yue06,2018_MindemarkLaceyEtAl}. In this regard, molecular modelling~\cite{zhang23} can be rather useful to disentangle different factors that influence the mechanical properties of ionic conductive polymers and to extract design principles.

In contrast to ionic transport properties (e.g. transference number) where much has been understood recently with the help of molecular modelling~\cite{2019_FranceLanordGrossman, fong19, zhang2020, Shen.2020, bollinger20, Gudla2020, Gudla2021, 2022.Halat, Shao2022, Shao2023}, the mechanical properties of ionic conductive polymers are less studied~\cite{mogurampelly16, verners16, sampath17, Verners18, sampath18}, in particular, at atomistic scale. Therefore, in this work,   we used a model ionic conductive polymer system, i.e.,
poly(ethylene oxide)-LiTFSI, and all-atom molecular dynamics (MD) simulations to study the impact of salt concentrations on
mechanical properties, including different types of elastic moduli and the visoelasticity. It is found that all-atom force fields popularly used in the studies of ion transport in polymer systems can reproduce quite well the experimental results of Young's modulus, bulk modulus and viscosity. Despite of the general trade-off between the transport property and the mechanical property as bounded by the Maxwell relation, we are able to identified an intermediate salt concentration at which the system possess both high ionic conductivity and high Young’s modulus. Regarding the self-healing capability, we show that the elastic restoration time is correlated with the Young’s modulus in a non-linear manner, which is interesting for further investigation. 

In the following, we first present the theory of elasticity and viscoelasticity as well as the non-equilibrium and equilibrium MD methods used to investigate these mechanical properties. It follows with the results of salt effects on both elastic moduli and relaxation modulus. Then, we present our attempt to identify an optimal salt concentration and to quantify the self-healing capability. At the last, a conclusion of this study and a perspective for future works are also provided.

\section{Theory and Methods}

%\begin{itemize}
%\item Add simulation protocol details
%\item MD parameters like timestep, barostat and thermostat details
%\item equilibration and production times
%\end{itemize}
\subsection{Elastic moduli from non-equilibrium MD}

According to Hooke's law for elasticity\cite{Brinson2015}, the 6$\times$6 elastic
constant matrix C is determined by the partial derivatives of the stress tensor, $\sigma_{ij}$, with respect to the deformation or strain $\varepsilon_{kl}$ is,
\begin{equation}
    C_{\alpha\beta}=\dfrac{\partial\sigma_{ij}}{\partial\varepsilon_{kl}}
\end{equation}
where $\{i,j,k,l\} \in \{x,y,z\}$ and $\{\alpha, \beta\} \in \{1,2,3\}$.
With the Voigt notation $xx \rightarrow 1$, $yy \rightarrow 2$, $zz \rightarrow 3$, $yz \rightarrow 4$, $xz \rightarrow 5$ and $xy \rightarrow 6$, the stiffness matrix $\mathbf{C}$ involving 21 unique elements can be written as follows
\[\boldsymbol{C} =
\begin{bmatrix}
    C_{11} & C_{12} & C_{13} & C_{14} & C_{15} & C_{16} \\
    C_{21} & C_{22} & C_{23} & C_{24} & C_{25} & C_{26} \\
    C_{31} & C_{32} & C_{33} & C_{34} & C_{35} & C_{36} \\
    C_{41} & C_{42} & C_{43} & C_{44} & C_{45} & C_{46} \\
    C_{51} & C_{52} & C_{53} & C_{54} & C_{55} & C_{56} \\
    C_{61} & C_{62} & C_{63} & C_{64} & C_{65} & C_{66} \\
\end{bmatrix}
=
\begin{bmatrix}
    \frac{\sigma_{xx}}{\varepsilon_{xx}} & \frac{\sigma_{xx}}{\varepsilon_{yy}} & \frac{\sigma_{xx}}{\varepsilon_{zz}} & \frac{\sigma_{xx}}{\varepsilon_{yz}} & \frac{\sigma_{xx}}{\varepsilon_{xz}} & \frac{\sigma_{xx}}{\varepsilon_{xy}} \\ 
    \frac{\sigma_{yy}}{\varepsilon_{xx}} & \frac{\sigma_{yy}}{\varepsilon_{yy}} & \frac{\sigma_{yy}}{\varepsilon_{zz}} & \frac{\sigma_{yy}}{\varepsilon_{yz}} & \frac{\sigma_{yy}}{\varepsilon_{xz}} & \frac{\sigma_{yy}}{\varepsilon_{xy}} \\ 
    \frac{\sigma_{zz}}{\varepsilon_{xx}} & \frac{\sigma_{zz}}{\varepsilon_{yy}} & \frac{\sigma_{zz}}{\varepsilon_{zz}} & \frac{\sigma_{zz}}{\varepsilon_{yz}} & \frac{\sigma_{zz}}{\varepsilon_{xz}} & \frac{\sigma_{zz}}{\varepsilon_{xy}} \\ 
    \frac{\sigma_{yz}}{\varepsilon_{xx}} & \frac{\sigma_{yz}}{\varepsilon_{yy}} & \frac{\sigma_{yz}}{\varepsilon_{zz}} & \frac{\sigma_{yz}}{\varepsilon_{yz}} & \frac{\sigma_{yz}}{\varepsilon_{xz}} & \frac{\sigma_{yz}}{\varepsilon_{xy}} \\ 
    \frac{\sigma_{xz}}{\varepsilon_{xx}} & \frac{\sigma_{xz}}{\varepsilon_{yy}} & \frac{\sigma_{xz}}{\varepsilon_{zz}} & \frac{\sigma_{xz}}{\varepsilon_{yz}} & \frac{\sigma_{xz}}{\varepsilon_{xz}} & \frac{\sigma_{xz}}{\varepsilon_{xy}} \\ 
    \frac{\sigma_{xy}}{\varepsilon_{xx}} & \frac{\sigma_{xy}}{\varepsilon_{yy}} & \frac{\sigma_{xy}}{\varepsilon_{zz}} & \frac{\sigma_{xy}}{\varepsilon_{yz}} & \frac{\sigma_{xy}}{\varepsilon_{xz}} & \frac{\sigma_{xy}}{\varepsilon_{xy}} 
\end{bmatrix}.
\]

For an isotropic and cubic system, $\boldsymbol{C}$ is only dependent on two variable $\lambda$ and $\mu$ called Lam\'{e}'s constants,\cite{Suter2002}
\[\boldsymbol{C} \equiv
\begin{bmatrix}
    \lambda+2\mu & \lambda & \lambda & 0 & 0 & 0 \\
    \lambda & \lambda+2\mu & \lambda & 0 & 0 & 0 \\
    \lambda & \lambda & \lambda+2\mu & 0 & 0 & 0 \\
    0 & 0 & 0 & \mu & 0 & 0 \\
    0 & 0 & 0 & 0 & \mu & 0 \\
    0 & 0 & 0 & 0 & 0 & \mu \\
\end{bmatrix}
.
\]
A least square procedure can be used to obtain Lam\'{e}'s constants according to equations \ref{eq:mul}-\ref{eq:ccmat},\cite{Suter2002} 
\begin{eqnarray}
    \mu &=& \dfrac{4a-2b+3c}{33}; \lambda = \dfrac{2a+c-15\mu}{6} \\
    \label{eq:mul}
    a &=& C_{11}+C_{22}+C_{33} \\
    \label{eq:acmat}
    b &=& C_{12}+C_{21}+C_{23}+C_{32}+C_{13}+C_{31} \\
    \label{eq:bcmat}
    c &=& C_{44}+C_{55}+C_{66} 
    \label{eq:ccmat}
\end{eqnarray}

Further, they can be used to calculate elastic moduli such as Young's modulus ($E$), shear modulus ($G$), bulk modulus ($B$) and Poisson's ratio ($\nu$) according to the following equation.
\begin{eqnarray}
    E=\frac{\mu(3\lambda+2\mu)}{\lambda + \mu} ;\ \
    G=\mu ;\ \
    B=\lambda+\frac{2}{3}\mu ;\ \
    \nu=\dfrac{\lambda}{2(\lambda+\mu)}
    \label{eq:moduli}
\end{eqnarray}

\begin{figure}[ht]
  \includegraphics[width=1\textwidth]{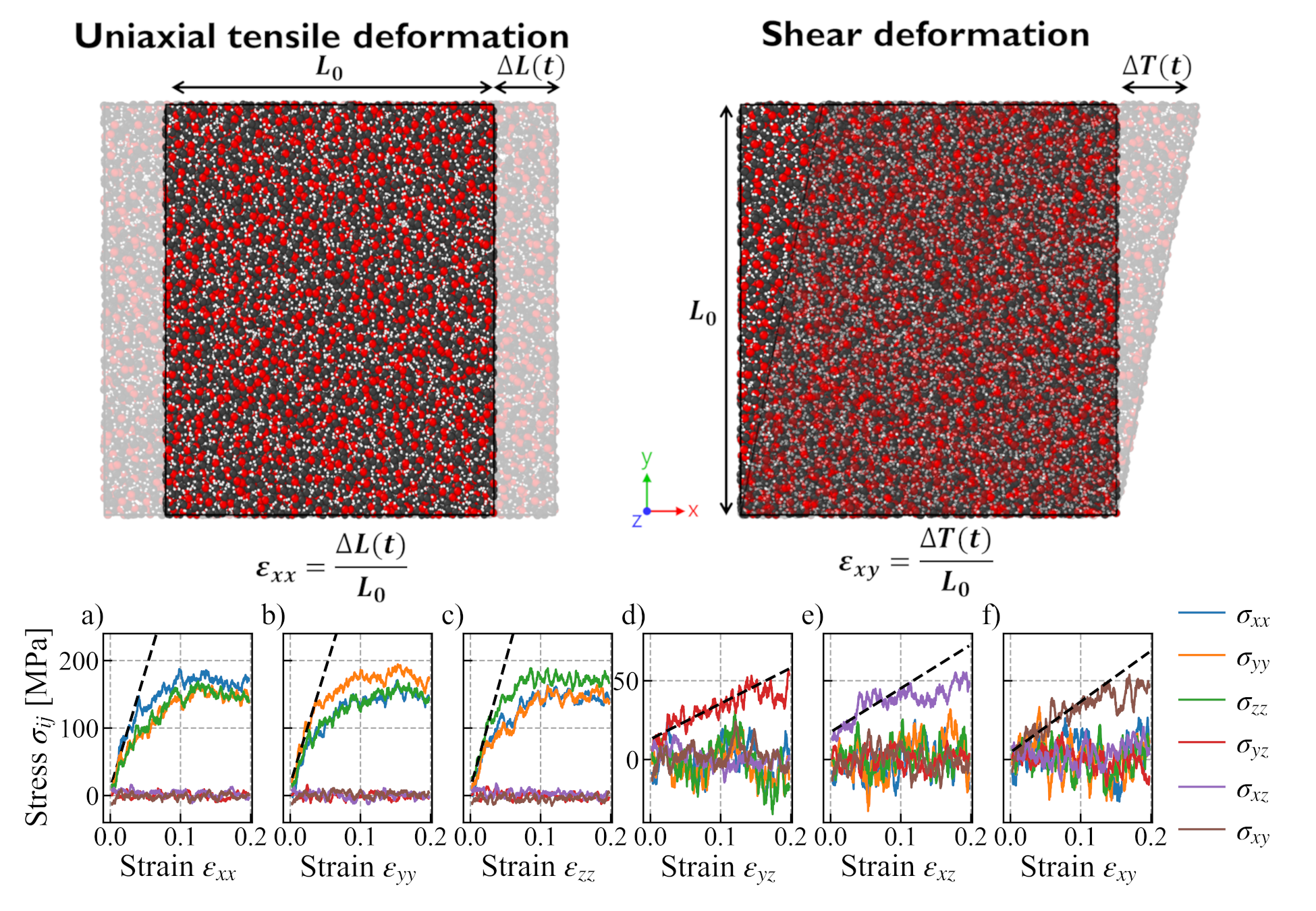}
  \caption{(top) Initial and final configurations of uniaxial tensile deformation and shear deformation simulations; (bottom a-f) six stress tensors plotted against strain in six non-equilibrium deformation simulations for the neat PEO system at a strain rate $\dot{\varepsilon}$ of $5\times10^9$ s$^{-1}$ and $T-T_\textrm{g} \approx 120$ K. Black dashed lines are linear fits to obtain $C_{\alpha\beta}$ values. }
  \label{fgr:sims_cmat}
\end{figure}

A series of deformations, i.e. uniaxial tensile deformation and shear deformation (see Fig. \ref{fgr:sims_cmat} top), were applied to the periodic simulation cell in order to estimate values of matrix element $C_{\alpha\beta}$\cite{Clancy2009}. For the uniaxial tensile deformation simulation, a strain was applied to the $x$ direction and the remaining two dimensions ($yz$) were unchanged. This was repeated for the $y$ and $z$ directions, while keeping the remaining ($xz$ and $xy$, respectively) dimensions ﬁxed. Likewise, for three shear deformation simulations where shear strains were applied to $yz$, $xz$ and $xy$ planes. All deformations were in the positive direction, the strain was applied in a continuous fashion at every time step at a constant rate and six different strain rates varying from $10^8$ - $5\times10^{10}~s^{-1}$ were considered in each case. For example, as shown in Fig.~\ref{fgr:sims_cmat} bottom, in the case of the deformation of the $x$ direction, the $\sigma_{xx}$ was plotted as a function of the strain $\varepsilon_{xx}$ up to 20\%, and the slope of this curve in the elastic regime over 5\% of strain was obtained. Similarly, plots of $\sigma_{yy}$ and $\sigma_{zz}$ versus the strain $\varepsilon_{xx}$ and the corresponding slopes were obtained. These results were placed in the 1st column of the $\mathbf{C}$ matrix. The ﬁrst three columns of the $\mathbf{C}$ matrix were therefore obtained from the three independent tensile deformation simulations in $x$, $y$ and $z$ directions. Instead, the last three columns were obtained from the three independent shear deformation simulations in $yz$, $xz$ and $xy$ planes.

\subsection{Viscoelastic properties from equilibrium MD}
From the Green-Kubo relation, the relaxation modulus $G(t)$ of the system can be obtained from the autocorrelation functions of off-diagonal stress component $\langle\sigma_{xy}(t)\sigma_{xy}(0)\rangle$ recorded during the equilibrium MD simulations.\cite{David2019} In the isotropic systems, $G(t)$ can be obtained
by averaging autocorrelations over the symmetrized traceless stress tensor ($\tau_{ij}$) components according to Eq. \ref{eq:gt_eq} to reduce the statistical error~\cite{2012.Liu20a}. 
\begin{eqnarray}
    G(t) &=& \frac{V}{10k_\textrm{B}T}\sum_i\sum_j\langle\tau_{ij}(t)\tau_{ij}(0)\rangle \\
    \label{eq:gt_eq} 
    \tau_{ij}(t) &=& \frac{1}{2}[\sigma_{ij}(t)+\sigma_{ji}(t)]-\frac{\delta_{ij}}{3}\sum_k \sigma_{kk}(t)
    \label{eq:tra_str}
\end{eqnarray}
where $V$ is the volume of the system, $T$ is the temperature and $k_\textrm{B}$ is the Boltzmann constant. 

The storage modulus $G^{\prime}(\omega)$ and the loss modulus $G^{\prime\prime}(\omega)$ can be computed from the in-phase (real) and out-of-phase (imaginary) components of the relaxation modulus in the frequency domain (Eq. \ref{eq:g11_gt}).
\begin{eqnarray}
    G^{\prime}(\omega) = \omega \int_0^\infty G(t)\sin(\omega t)dt \ ; \
    G^{\prime\prime}(\omega) = \omega \int_0^\infty G(t)\cos(\omega t)dt 
    \label{eq:g11_gt}
\end{eqnarray}

To reduce the fluctuations in the stress time-autocorrelation functions, the multi-tau correlator method\cite{Ramirez2010} was used to calculate $G(t)$ by implementing the python code ``multipletau" \cite{multau}. The numerical evaluations of $G^{\prime}(\omega)$ and $G^{\prime\prime}(\omega)$ were carried by following the method proposed by Adeyemi $et.\ al$\cite{Adeyemi2022,Adeyemi2022a}.

Then, the modulus of the frequency-dependent viscosity $|\eta^*(\omega)|$ can then be estimated from the storage and loss moduli using the following expression:
\begin{equation}
    |\eta^*(\omega)|=\frac{\sqrt{G{'}(\omega)^2+G{''(\omega})^2}}{\omega}.
    \label{eq:eta_star}
\end{equation}

Taking the zero frequency limit, one obtains the equilibrium viscosity $\eta_0$:
\begin{equation}
    \eta = \lim_{\omega \rightarrow 0}|\eta^*(\omega)|
    \label{eq:eta_0} = \int_{0}^\infty dt G(t)
\end{equation}

Equivalently, the equilibrium viscosity can also be obtained from the the following expression:
\begin{equation}
\label{eq:eta_0_2}
    \eta = \lim_{\omega \rightarrow 0}\frac{G''(\omega)}{\omega}
\end{equation}

\subsection{MD simulations of PEO-LiTFSI systems}

The GAFF force field parameters\cite{Wang2004} and simulation protocol for PEO-LiTFSI systems (25 monomer units in each PEO chain with the molecular weight of 1.11 kg/mol) at six different concentrations can be found in our previous studies~\cite{Gudla2020, Gudla2021, Shao2022, Shao2023}. In this study, all MD simulation were carried out using LAMMPS \cite{LAMMPS} instead of GROMACS\cite{Abraham2015} for the convenience of computing mechanical properties. To ensure the consistency with our previous studies, glass transition temperatures and Nernst-Einstein ionic conductivity ($\sigma_\textrm{NE}$) at different concentrations  were computed using two codes but the same force field parameters and compared (see Fig.S1 in the Supporting Information). Non-equilibrium MD simulation were performed for 1-10 ns depending on the strain rate and equilibrium MD simulations were carried out for 500-600 ns. For non-equilibrium MD simulations, we have applied Nos\'{e}-Hoover thermostat\cite{Evans85} with SLLOD equations of motion\cite{Evans84} at $T-T_\textrm{g} \approx 60$ K (room temperature) and $T-T_\textrm{g} \approx 120$ K (about 430 K) respectively. For 
 equilibrium MD simulations, Nos\'{e}-Hoover thermostat\cite{Evans85} and barostat\cite{Noose83} were applied at $T-T_\textrm{g} \approx 120$ K and 1 bar.

\section{Results and Discussion}
\subsection{Effects of strain rate and salt concentration on elastic properties}
\begin{figure}[ht]
  \includegraphics[width=1\textwidth]{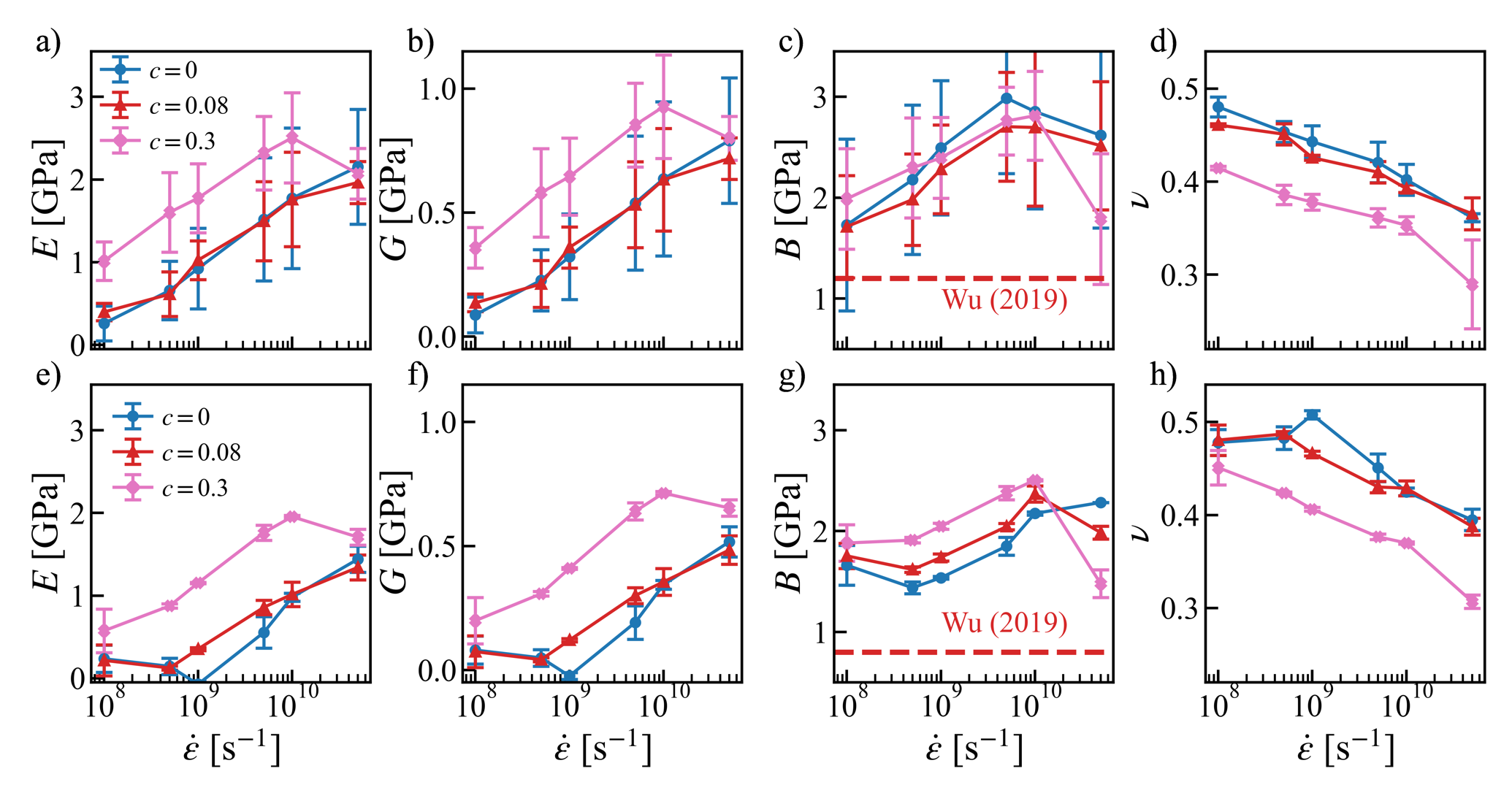}
  \caption{a,e) Young's modulus $E$; b,f) shear modulus $G$; c,g) bulk modulus $B$ and d,h) Poisson's ratio $\nu$ as a function of strain rate ($\dot{\varepsilon}$) for the neat PEO and the PEO-LiTFSI systems at two different concentrations ($c=0.02,0.3$) and at $T-T_\textrm{g}\approx60$ K (a,b,c,d) and $120$K (e,f,g,h). Red dashed line: The experimental bulk modulus for the neat PEO system at the respective temperatures from Ref.~\citenum{Wu2019}.  }
  \label{fgr:mod_cmat}
\end{figure}

The elastic moduli and Poisson's ratio calculated from Eq. \ref{eq:moduli} as a function of strain rate at two different effective temperatures ($T-T_\textrm{g}$) and three salt concentrations ($c$) are shown in Fig. \ref{fgr:mod_cmat}. The error bars were estimated using the standard deviation from  simulations with 5 different initial configurations. As shown in Fig. \ref{fgr:mod_cmat}, both Young's modulus $E$ and shear modulus $G$ were found to increase with the strain rate, as the polymer chains have less time to response to the strain which makes the chains look stiffer. The opposite was seen for the Poisson's ratio, which approaches a perfect incompressible rubber state at a lower strain rate~\cite{Greaves2011}. In contrast, the bulk modulus $B$ seems to be insensitive to the strain rate. This explains why the simulation results agree quite well with the experimental bulk modulus~\cite{Wu2019}, despite of the order of magnitude difference in the strain rate. Since the Lam\'e constant $\lambda$ is the dominating contribution to $B$ (see Eq.~\ref{eq:moduli}, Figs. S3b,d), this also explains why the Poisson's ratio has an opposite strain-rate dependence as compared to $E$ and $G$. 

\begin{figure}[ht]
  \includegraphics[width=1\textwidth]{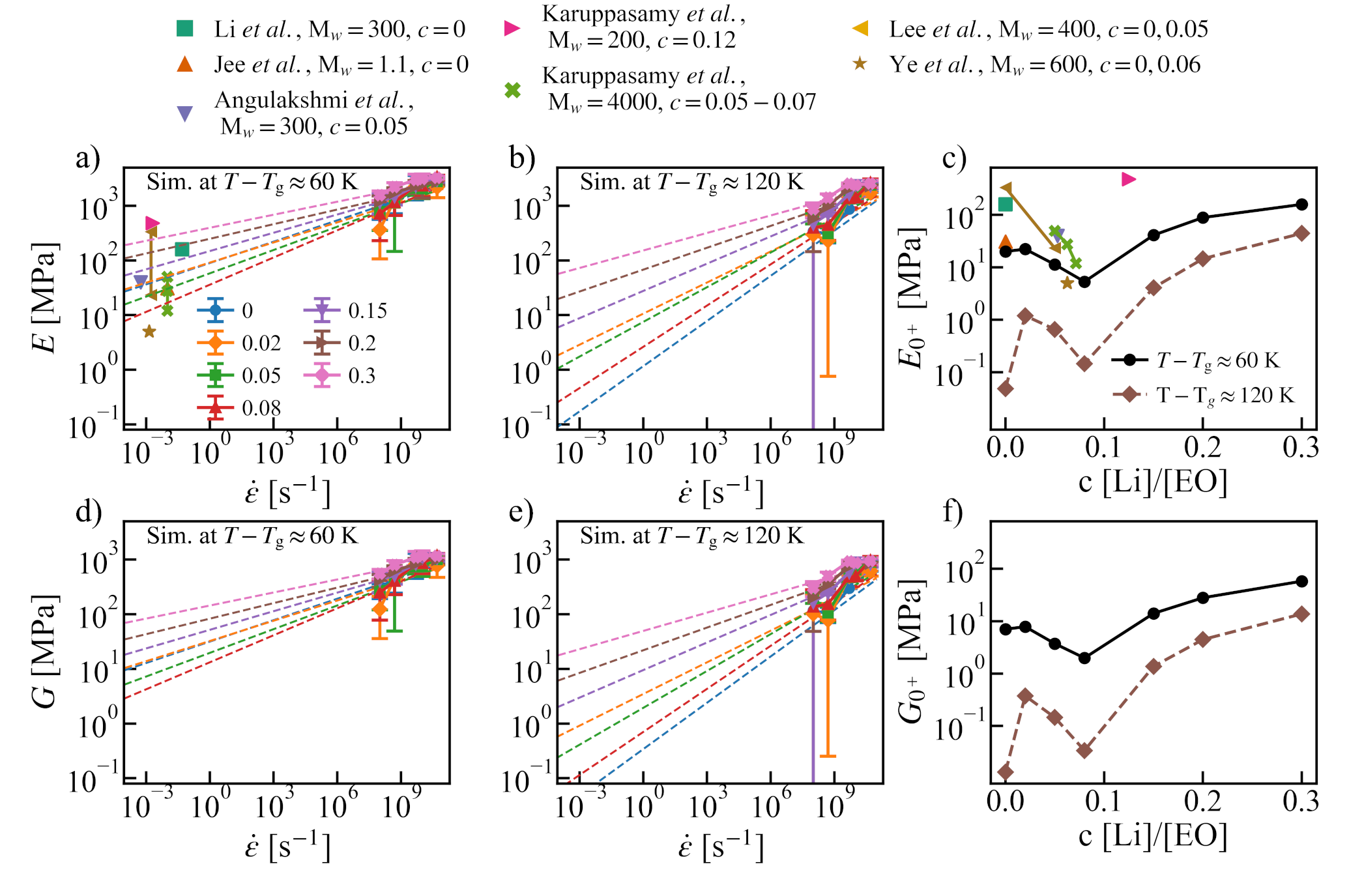}
  \caption{ a, b) Young's modulus $E$ and d, e) shear modulus $G$ as a function of strain rate ($\dot{\varepsilon}$) for the neat PEO and the PEO-LiTFSI systems at temperatures $T-T_\textrm{g}\approx60$ K and $120$ K from non-equilibrium MD simulations. c) The near zero strain-rate Young's modulus $E_{0^+}$ and f) near zero strain-rate shear modulus $G_{0^+}$ as a function of salt concentration. Dashed lines are linear fits to calculate Young's modulus at near zero strain rate $E_{0^+}$. The experimental Young's modulus for PEO based electrolytes at room temperature~\cite{Li2020,Jee2013,Angulakshmi2014,Karuppasamy2016,Karuppasamy2017,Lee2019,Ye2015} were also plotted for comparison and summarized in the Supporting Information.} 
  \label{fgr:ey_sim-exp}
\end{figure}

The strain rates used in the simulations are 8–10 orders of magnitude higher in comparison to the experimental strain rates ($10^{-4} - 10^{-2}$ s$^{-1}$). Therefore, in order to compare with experimental Young's modulus, the extrapolation was used. In Figs. \ref{fgr:ey_sim-exp}a, b, d, and e, the $E$ and $G$ were plotted as function of strain rates and linear fittings in the log-log scale were carried out to estimate the near zero strain-rate Young's modulus $E_{0^+}$ i.e., $E$ at a strain rate ( $10^{-2}$ s$^{-1}$) similar to the experimental one. As shown Fig. \ref{fgr:ey_sim-exp}a, the extrapolations (dashed line) to the low strain rate are in very good agreement with the experimental values for PEO-based electrolytes in the same effective temperatures $T-T_\textrm{g}\approx 60$K.\cite{Li2020,Jee2013,Angulakshmi2014,Karuppasamy2016,Karuppasamy2017,Lee2019,Ye2015} As shown in  Figs. \ref{fgr:mod_cmat} and \ref{fgr:ey_sim-exp}b and \ref{fgr:ey_sim-exp}e, it is clear that increasing the effective temperature will reduce both the Young's modulus and the shear modulus. 

The salt concentration has a non-monotonic effect on both elastic moduli and Poisson's ratio, which can be already seen in Fig.~\ref{fgr:mod_cmat}. At high salt concentration $c=0.3$, both $E$ and $G$ become much larger as compared to the neat PEO system. However, at low and moderate concentrations, the effect can be the opposite and the increase in temperature may further convolute the situation. This can be clearly seen in Figs.~\ref{fgr:ey_sim-exp}c and \ref{fgr:ey_sim-exp}f, which shows the salt concentration-dependence of $E_{0^+}$ and $G_{0^+}$. Simulations results agree with the experimental trend that the elastic moduli decreases with a moderate increment in salt concentration (Figs.~\ref{fgr:ey_sim-exp}c). However, a further increase in the salt concentration beyond $c=0.08$ leads to enhanced elastic moduli instead as predicted by simulations.

\subsection{Salt effects on relaxation modulus and viscosity}
\begin{figure}[ht]
  \includegraphics[width=0.9\textwidth]{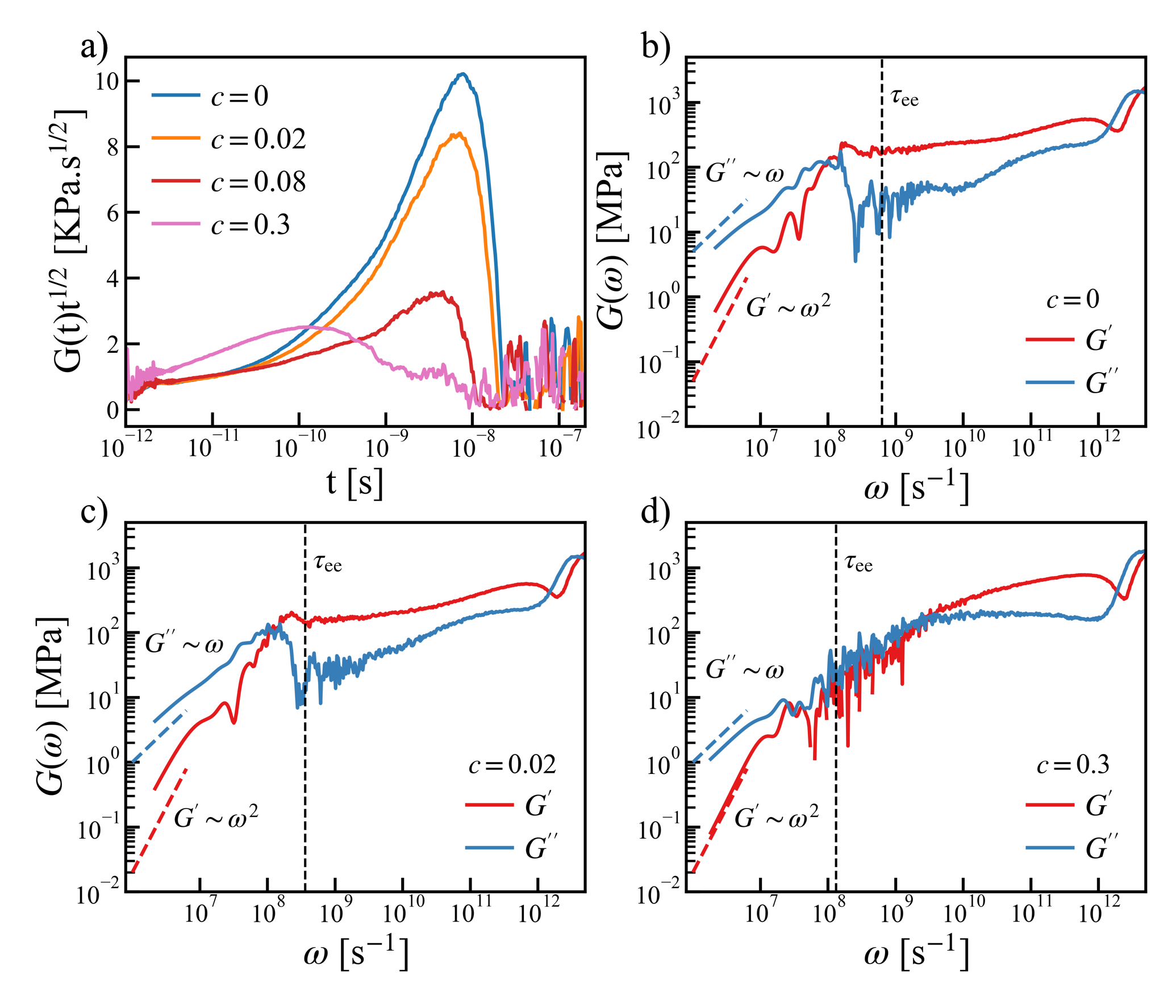}
  \caption{a) The relaxation modulus $G(t)$ scaled with $t^{1/2}$ for PEO and PEO-LiTFSI at different concentrations ($c=0.02, 0.08, 0.3$). b, c, d) The storage modulus $G^{\prime}(\omega)$ and the loss modulus $G^{\prime\prime}(\omega)$ for the neat PEO and PEO-LiTFSI systems at $c=0.02$, $0.3$ at $T-T_\textrm{g}\approx120$ K. The blue and red dashed lines corresponds to the relations $G{''} \sim\omega$ and $G{'} \sim\omega^2$. The vertical dashed lines indicate the end-to-end relaxation time $\tau_{ee}$ (see Supporting Information).}
  \label{fgr:gt_calc}
\end{figure}

The shear stress relaxation moduli $G(t)$ calculated from EMD simulations for different salt concentrations are shown in Fig. \ref{fgr:gt_calc}a.  Likthman $et.\ al$\cite{Likhtman2007} has identified four time scales to the stress relaxation modulus by using a simple bead-spring model of polymer melt, namely i) the oscillatory behaviours at short time arises due to bond length relaxations; ii) the colloidal or
glassy mode due to collisions between atoms; iii) the Rouse dynamics i.e. polymer relaxation according to the Rouse theory $G(t)\sim t^{-1/2}$; iv) the polymer entanglement. Given the rigid bond model and low molecular weight systems used in this study, it is natural for us to focus on identifying the Rouse dynamics. If the system follows the Rouse dynamics, then  the product $G(t)t^{1/2}$ would equal to a constant.  As shown in Fig. \ref{fgr:gt_calc}a, with the addition of salt, these dynamics seem to gradually deviate from the Rouse dynamics. The system with the highest salt concentration ($c=0.3$) shows the largest deviation from the Rouse theory. It is also interesting to note that the large peak shown in Fig. \ref{fgr:gt_calc}a resembles the entanglement behaviour of high molecular weight systems revealed in Ref.~\citenum{Likhtman2007}, and we will come back to this point in the paragraph below. 

\begin{figure}[ht]
  \includegraphics[width=1\textwidth]{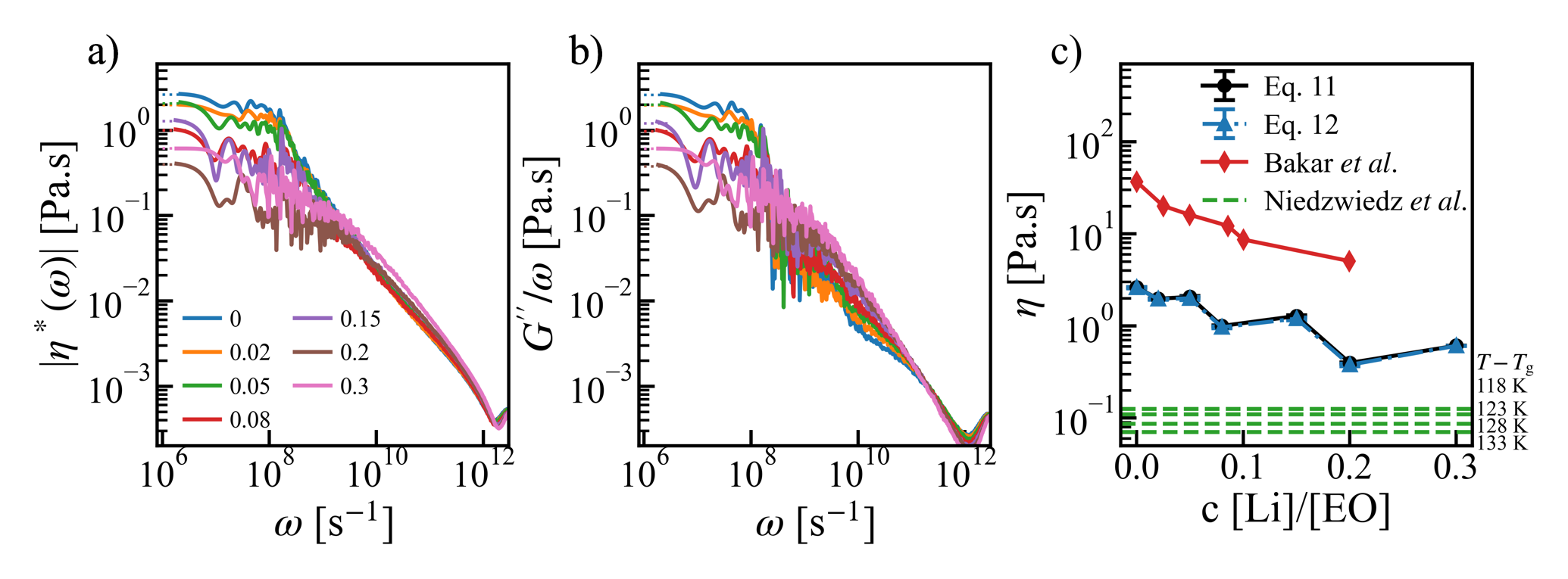}
  \caption{a) The modulus of the frequency-dependent viscosity $|\eta^*(\omega)|$ and b) $G''(\omega)/\omega$ for the neat PEO and PEO-LiTFSI systems at different concentrations and at temperature $\mathrm{T-T}_g\approx120$ K. b) Comparing the equilibrium viscosity $\eta$ computed from Eq.~11 and Eq.~12 with experimental values~\cite{Bakar2023,Niedzwiedz} for PEO-LiTFSI systems as a function of salt concentration. }
  \label{fgr:eta_calc_exp}
\end{figure}

The storage modulus $G^{\prime}(\omega)$ and the loss modulus $G^{\prime\prime}(\omega)$ for the neat PEO ($c=0$) and two concentrations ($c=0.02$ and $c=0.3)$ were plotted in Fig. \ref{fgr:gt_calc} b, c, d. In all cases, a clear crossover from the solid-like behaviours $G^{\prime}(\omega)>G^{\prime\prime}(\omega)$ to the liquid-like behaviours $G^{\prime\prime}(\omega)>G^{\prime}(\omega)$ was observed.  In the low frequency range, the asymptotic behaviors of viscoelastic liquid \cite{Verdier2003} $G{''} \sim\omega$ and $G{'} \sim\omega^2$ were also evinced in our simulations. In addition, a second cross-over at high frequency were seen in all cases from simulations. It is worth noting that the linear rheology experiments are usually conducted at much lower frequency (longer time scale) and at higher molecular weights~\cite{Niedzwiedz, behbahani21} and the second cross-over between $G'$ and $G''$ signals the entangled polymer dynamics~\cite{Rubinstein}. Nevertheless, similar observations made here suggest that the viscoelastic properties from low molecular weight system and equilibrium MD simulations may emulate the realistic polymer dynamics at much higher molecular weight and longer time scale.  

 To make a further connection to experiment, we plotted the modulus of the complex viscosity in Fig. \ref{fgr:eta_calc_exp}a for different salt concentrations at $T-T_\textrm{g}=120 K$. As expected from Eq.~\ref{eq:eta_0}, we observed that at lower frequency the values tend to become a constant and can be used to estimate the equilibrium viscosity $\eta$. The same applies to the estimator based on Eq.~\ref{eq:eta_0_2} using the loss modulus $G''$ (see Fig. \ref{fgr:eta_calc_exp}b). The results of $\eta$ are shown in Fig.~\ref{fgr:eta_calc_exp}c. With an increase in concentration, $\eta$ tends to decrease with oscillations. A similar trend was also observed in the experiments \cite{Bakar2023} for PEO-LiTFSI systems (the red line in Fig.~\ref{fgr:eta_calc_exp}c). However, the viscosity values reported in experiment are 1 order of magnitude higher than the simulation results obtained here because of the difference in the molecular weight (20 kg/mol in experiment versus 1.1 kg/mol in simulation). Indeed, our results come closer to experimental reference\cite{Niedzwiedz} measured at similar molecular weight (green dashed lines in Fig.~\ref{fgr:eta_calc_exp}c).

 Before closing this section, it is worth noting that both the storage and loss modulus decrease with the increment in salt concentration (see Fig.~\ref{fgr:gt_calc}b, c, d), which is similar to that of the equilibrium viscosity $\eta$. This may appear in contradiction with the finding shown in Fig.~\ref{fgr:ey_sim-exp}f that the shear modulus $G_{0^+}$ increases with salt doping especially at high concentration. However, in the solid-like regime, i.e. $G^{\prime}(\omega)>G^{\prime\prime}(\omega)$ at high frequency, the magnitude of $G'$ indeed becomes larger by adding salts. Therefore, this contrast just reflects the opposite effects of salts on the shear modulus of ionic conductive polymers at different time scales. Furthermore, according to the Maxwell model, which is the combination of a Hookean solid and a Newtonian fluid, the shear stress relaxation time is determined by the ratio $\eta/G_{0^+}$. This means the shear stress relaxation time should decrease more rapidly with the increase of the salt concentration. Indeed, this is borne, as shown by the cross-over time $G'= G''$ at the low frequency in Fig.~\ref{fgr:gt_calc}.  

\subsection{Optimal salt concentration and self-healing capability}

As stated in the Introduction, searching ionic conductive polymers that satisfy the requirements for both transport and mechanical properties and demonstrate self-healing functionality is an emerging topic in battery field. Therefore, it would be interesting to address this point from our simulations. 

\begin{figure}[ht]
    \centering
    \includegraphics[width=1\textwidth]{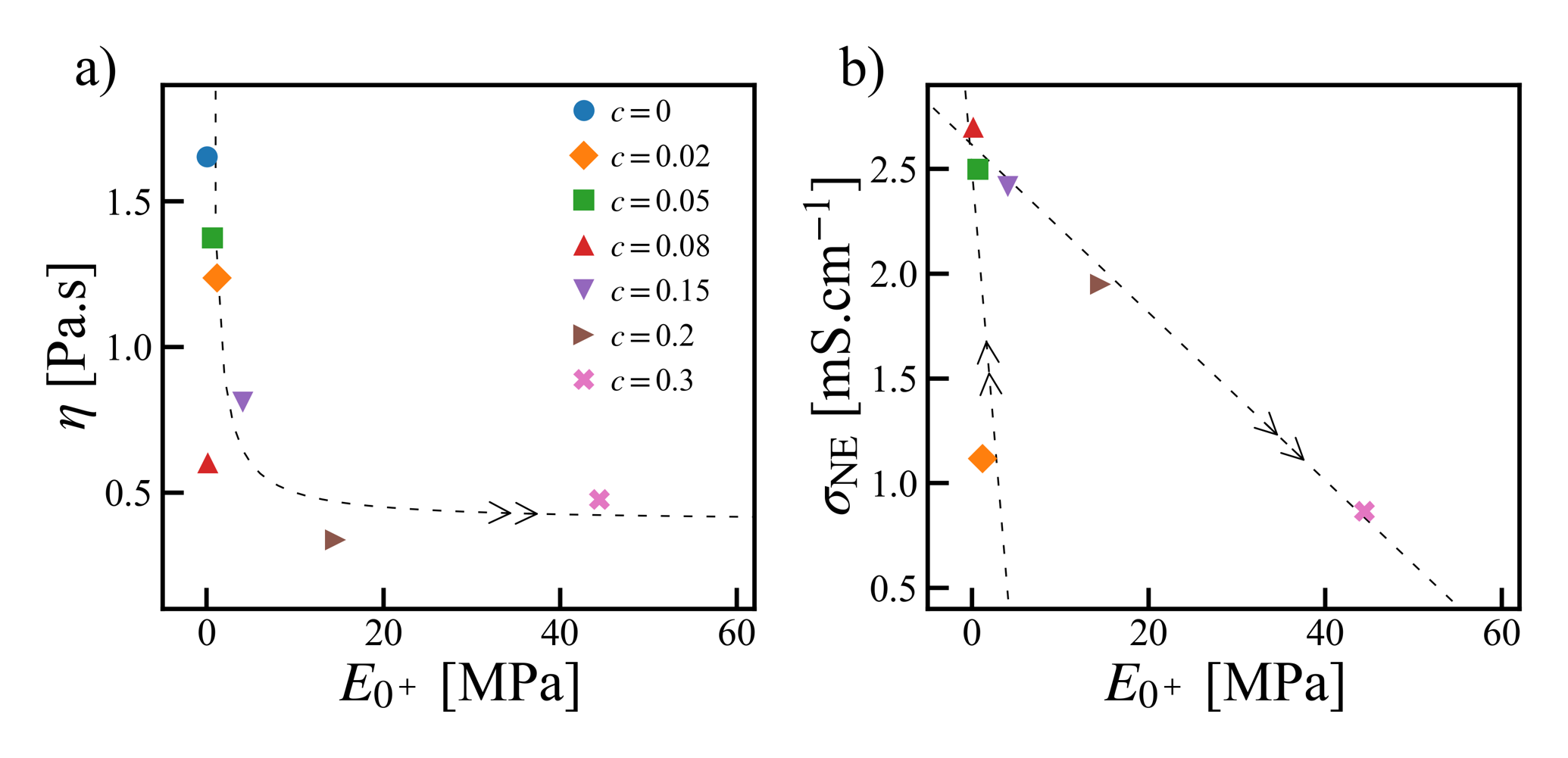}
    \caption{Correlations between the Young's modulus at near zero strain rate $E_{0^+}$ with (a) the equilibrium viscosity $\eta$  and (b) the Nernst-Einstein conductivity $\sigma_\textrm{NE}$  at $T-T_\textrm{g} \approx 120$ K. The dashed lines are guides to the eye. The arrow heads indicate the direction where the salt concentration increases.}
    \label{fig:E0_corr}
\end{figure}

As shown in Fig.~\ref{fig:E0_corr}a, the salt effects on the equilibrium viscosity $\eta$ and the near zero strain-rate Young's modulus $E_{0^+}$ are opposite, which makes these two quantities anti-correlated to each other. This means, there is a trade off between a good Newtonian fluid and a good Hookean solid, which is what the Maxwell relation implies~\cite{Angell:1998vw}. 

However, when making the correlation between the ionic conductivity $\sigma_\textrm{NE}$ and the Youngs' modulus $E_{0^+}$, the situation is more interesting. Despite that $\sigma_\textrm{NE}$ and $E_{0^+}$ is also anti-correlated in general, there exist two regimes. From the low to the intermediate concentrations,   $\sigma_\textrm{NE}$ goes up rapidly while $E_{0^+}$ slightly goes down; with a further increment in the concentration, $\sigma_\textrm{NE}$ goes down while $E_{0^+}$ goes up in a comparable degree. As a consequence, there is an intermediate salt concentration at $c = 0.2$ where the system processes both high ionic conductivity and high Young's modulus. It is worth noting that at a lower temperature, $\sigma_\textrm{NE}$ goes down while $E_{0^+}$ goes up. Therefore, an optimal salt concentration may be located where the requirements for both ionic conductivity and mechanical strength as mentioned in the Introduction can be achieved. 

\begin{figure}[ht]
    \centering
    \includegraphics[width=1\textwidth]{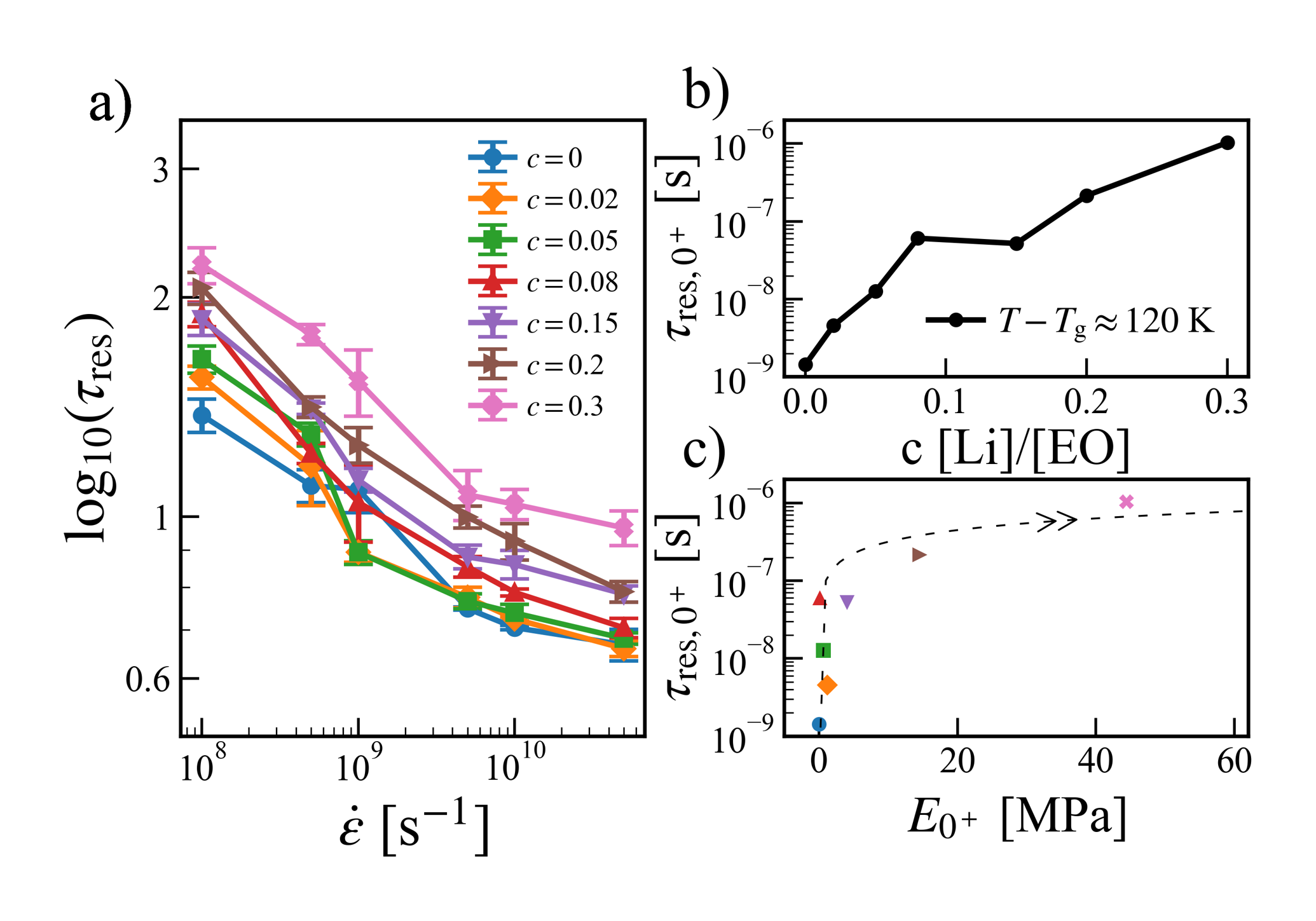}
    \caption{a) The elastic restoration time $\tau_\textrm{res}$ as a function of strain rate $\dot{\varepsilon}$ at different salt concentrations; b) The elastic restoration time at the near zero strain rate $\tau_{\textrm{res}, {0^+}}$ as a function of salt concentration ; c) The correlations between the elastic restoration time at the near zero strain rate $\tau_{\textrm{res}, {0^+}}$ and the near zero-strain rate Young's modulus $E_{0^+}$ at $T-T_\textrm{g} \approx 120$ K. The dashed line is guides to the eye and the arrow heads indicate the direction where the salt concentration increases.}
    \label{fig:tau_res}
\end{figure}

The final point that we want to address here is about the self-healing capability of ionic conductive polymers. Here, we define the self-healing capability as the elastic restoration time $\tau_\textrm{res}$ for the system to restore its equilibrium density after the expansion under a tensile strain (see the Supporting Information). As shown in Fig.~\ref{fig:tau_res}a, despite all simulations used to compute the restoration time $\tau_\textrm{res}$ started with the same expansion rate of 20\%, the resulting $\tau_\textrm{res}$ depends on the strain rate used in the generated these initial structures. This suggests that the self-healing capability depends on the history of how fast the deformation has taken place. Therefore, we used the elastic restoration time $\tau_{\textrm{res}, {0^+}}$ extrapolated to the near zero strain rate as a benchmark index.

As shown in Fig.~\ref{fig:tau_res}b, $\tau_{\textrm{res},{0^+}}$ increases with the salt concentration. This implies that $\tau_{\textrm{res}, {0^+}}$ correlate positively with the Young's modulus $E_{0^+}$, as seen in Fig.~\ref{fig:tau_res}c. In other words, ionic conductive polymers of the highest Young's modulus do not have the best self-healing capacity.  However, this correlation is non-linear, and the system with $c = 0.2$ shows a good balance between high Young's modulus and short elastic restoration time.

\section{Conclusions}
In this study, we have carried out all-atom MD simulations to investigate salt effects on the
mechanical properties of
poly(ethylene oxide)-LiTFSI as a model ionic conductive polymer system with both non-equilibrium and equilibrium methods. The focus has been on both the elastic moduli and the relaxation modulus. 

Regarding the elastic moduli, it is found that all-atom force fields commonly used in studying ion transport can
reproduce quite well the experimental results of Young’s modulus and bulk modulus. Further, we found that the Poisson’s ratio goes down by increasing the strain-rate while the opposite happens to the Young's modulus $E$ and shear modulus $G$.  We confirmed the experimental observation that in the low concentration regime, the Young's modulus becomes smaller by adding salts. However, our simulation also revealed that a further increasing of the salt concentration can enhance the Young's modulus instead. 

In terms of the relaxation modulus,  our MD simulations showed that the low molecular weight system and
equilibrium MD simulations may emulate the entanglement features of the relaxation modulus, which should only happen in principle to polymer systems at much higher molecular weight and longer time scale. Moreover, the computed viscosity $\eta$ is in a good agreement with experimental results at a comparable molecular weight and we confirmed the experimental observation of a decrement in viscosity with salt concentration. The same trend was also seen for both the storage modulus $G'$ and the loss modulus $G''$ at the low frequency regime from simulations. 

Besides comparing the results with experiments and studying the trends, we were able to identify an intermediate salt concentration $c=0.2$ at which the system possess both high ionic conductivity and high Young’s modulus. This intermediate salt concentration also leads to a short elastic restoration time, which can be relevant to the self-healing capacity of ionic conductive polymer. 

We expect that more follow-up studies will come out to relate the self-healing capacity of ionic conductive polymers to their mechanical properties with all-atom MD simulations. In particular, questions such as how to define self-healing capacities from MD simulations and how to relate them to measurable experimental quantities should be addressed. By making these efforts, we would be able to understand the molecular mechanisms of self-healing functionality and to extract design principles for novel polymer binder materials.

\begin{acknowledgement}

\end{acknowledgement}
This work was supported by the Knut and Alice Wallenberg Foundation (INTELiSTORE 139501042). The authors thank the funding from the Swedish National Strategic e-Science program eSSENCE, STandUP for Energy, and BASE (Batteries Sweden). The simulations were performed on the resources provided by the National Academic Infrastructure for Supercomputing in Sweden (NAISS) at PDC partially funded by the Swedish Research Council through Grant Agreement No. 2022-06725 and through the project access to the LUMI supercomputer, owned by the EuroHPC Joint Undertaking, hosted by CSC (Finland) and the LUMI consortium.

\begin{suppinfo}

Comparison between LAMMPS and GROMACS simulations with the same force field implementation; strain-rate dependence of Lam\'{e}'s constants $\mu$ and $\lambda$; summary of experimental references on the Youngs' modulus and viscosity; calculations of elastic restoration time $\tau_\textrm{res}$.

\end{suppinfo}

% \bibliography{References_HG}

\begin{mcitethebibliography}{63}
\providecommand*\natexlab[1]{#1}
\providecommand*\mciteSetBstSublistMode[1]{}
\providecommand*\mciteSetBstMaxWidthForm[2]{}
\providecommand*\mciteBstWouldAddEndPuncttrue
  {\def\EndOfBibitem{\unskip.}}
\providecommand*\mciteBstWouldAddEndPunctfalse
  {\let\EndOfBibitem\relax}
\providecommand*\mciteSetBstMidEndSepPunct[3]{}
\providecommand*\mciteSetBstSublistLabelBeginEnd[3]{}
\providecommand*\EndOfBibitem{}
\mciteSetBstSublistMode{f}
\mciteSetBstMaxWidthForm{subitem}{(\alph{mcitesubitemcount})}
\mciteSetBstSublistLabelBeginEnd
  {\mcitemaxwidthsubitemform\space}
  {\relax}
  {\relax}

\bibitem[Amici \latin{et~al.}(2022)Amici, Asinari, Ayerbe, Barboux,
  Bayle-Guillemaud, Behm, Berecibar, Berg, Bhowmik, Bodoardo, Castelli,
  Cekic-Lasko\~vic, Christensen, Clark, Diehm, Dominko, Fichtner, Franco,
  Grimaud, Guillet, Hahlin, Hartmann, Heiries, Hermansson, Heuer, Jana,
  Jabbour, Kallo, Latz, Lorrmann, Løvvik\, Lyonnard, Meeus, Paillard, Perraud,
  Placke, Punckt, Raccurt, Ruhland, Sheridan, Stein, Tarascon, Trapp, Weil,
  Wenzel, Winter, Wolf, and Edström]{bat2030}
Amici,~J. \latin{et~al.}  A Roadmap for Transforming Research to Invent the
  Batteries of the Future Designed within the European Large Scale Res\ earch
  Initiative BATTERY 2030+. \emph{Adv. Energy Mater.} \textbf{2022}, \emph{12},
  2102785\relax
\mciteBstWouldAddEndPuncttrue
\mciteSetBstMidEndSepPunct{\mcitedefaultmidpunct}
{\mcitedefaultendpunct}{\mcitedefaultseppunct}\relax
\EndOfBibitem
\bibitem[Narayan \latin{et~al.}(2021)Narayan, Laberty-Robert, Pelta, Tarascon,
  and Dominko]{Narayan.2021}
Narayan,~R.; Laberty-Robert,~C.; Pelta,~J.; Tarascon,~J.-M.; Dominko,~R.
  {Self-Healing: An Emerging Technology for Next-Generation Smart Batteries}.
  \emph{Adv. Energy Mater.} \textbf{2021}, 2102652\relax
\mciteBstWouldAddEndPuncttrue
\mciteSetBstMidEndSepPunct{\mcitedefaultmidpunct}
{\mcitedefaultendpunct}{\mcitedefaultseppunct}\relax
\EndOfBibitem
\bibitem[Liu \latin{et~al.}(2017)Liu, Tai, Zhou, Sencadas, Zhang, Zhang,
  Konstan\~tinov, Guo, and Liu]{Liu.2017}
Liu,~Y.; Tai,~Z.; Zhou,~T.; Sencadas,~V.; Zhang,~J.; Zhang,~L.;
  Konstan\~tinov,~K.; Guo,~Z.; Liu,~H.~K. {An All-Integrated Anode via
  Interlinked Chemical Bonding between Double-Shelled–Yolk-Structured Silicon
  and\ Binder for Lithium-Ion Batteries}. \emph{Adv. Mater.} \textbf{2017},
  \emph{29}, 1703028\relax
\mciteBstWouldAddEndPuncttrue
\mciteSetBstMidEndSepPunct{\mcitedefaultmidpunct}
{\mcitedefaultendpunct}{\mcitedefaultseppunct}\relax
\EndOfBibitem
\bibitem[Yang \latin{et~al.}(2021)Yang, Zhang, Li, Takenaka, Liu, Zou, Chen,
  Du, Hong, Shang, Nakamura, Cai, Lan, Zheng, and Yamada]{Yang.2021dnr}
Yang,~S.; Zhang,~Y.; Li,~Z.; Takenaka,~N.; Liu,~Y.; Zou,~H.; Chen,~W.~n.;
  Du,~M.; Hong,~X.-J.; Shang,~R.; Nakamura,~E.; Cai,~Y.-P.; Lan,~Y.-Q.;
  Zheng,~Y.,~Qifeng a\ nd~Yamada; Yamada,~A. {Rational Electrolyte Design to
  Form Inorganic–Polymeric Interphase on Silicon-Based Anodes}. \emph{ACS
  Energy Lett.} \textbf{2021}, \emph{6}, 1811--1820\relax
\mciteBstWouldAddEndPuncttrue
\mciteSetBstMidEndSepPunct{\mcitedefaultmidpunct}
{\mcitedefaultendpunct}{\mcitedefaultseppunct}\relax
\EndOfBibitem
\bibitem[Zhou \latin{et~al.}(2022)Zhou, Li, Han, Li, He, and Wang]{Zhou22}
Zhou,~Y.; Li,~L.; Han,~Z.; Li,~Q.; He,~J.; Wang,~Q. {Self-Healing Polymers for
  Electronics and Energy Devices}. \emph{Chem. Rev.} \textbf{2022}, \emph{123},
  558--612\relax
\mciteBstWouldAddEndPuncttrue
\mciteSetBstMidEndSepPunct{\mcitedefaultmidpunct}
{\mcitedefaultendpunct}{\mcitedefaultseppunct}\relax
\EndOfBibitem
\bibitem[Chen \latin{et~al.}(2019)Chen, Li, Luo, Shi, Ma, Zhang, Boukhvalov,
  and L\~uo]{chen19}
Chen,~J.; Li,~F.; Luo,~Y.; Shi,~Y.; Ma,~X.; Zhang,~M.; Boukhvalov,~D.~W.;
  L\~uo,~Z. {A self-healing elastomer based on an intrinsic non-covalent
  cross-linking mechanism}. \emph{J. Mater. Chem. A} \textbf{2019}, \emph{7},
  15207--15214\relax
\mciteBstWouldAddEndPuncttrue
\mciteSetBstMidEndSepPunct{\mcitedefaultmidpunct}
{\mcitedefaultendpunct}{\mcitedefaultseppunct}\relax
\EndOfBibitem
\bibitem[Mi \latin{et~al.}(2019)Mi, Yang, Li, Zhuang, Chen, Li, and
  Zhang]{Mi.2019}
Mi,~H.; Yang,~X.; Li,~F.; Zhuang,~X.; Chen,~C.; Li,~Y.; Zhang,~P. {Self-healing
  silicon-sodium alginate-polyaniline composites originated from the
  enhancement hydrogen bonding for lith\ ium-ion battery: A combined simulation
  and experiment study}. \emph{J. Power Sources} \textbf{2019}, \emph{412},
  749--758\relax
\mciteBstWouldAddEndPuncttrue
\mciteSetBstMidEndSepPunct{\mcitedefaultmidpunct}
{\mcitedefaultendpunct}{\mcitedefaultseppunct}\relax
\EndOfBibitem
\bibitem[Xie \latin{et~al.}(2021)Xie, Hu, Li, and Zhang]{Xie2021}
Xie,~Z.; Hu,~B.-L.; Li,~R.-W.; Zhang,~Q. {Hydrogen Bonding in Self-Healing
  Elastomers}. \emph{ACS Omega} \textbf{2021}, \emph{6}, 9319--9333\relax
\mciteBstWouldAddEndPuncttrue
\mciteSetBstMidEndSepPunct{\mcitedefaultmidpunct}
{\mcitedefaultendpunct}{\mcitedefaultseppunct}\relax
\EndOfBibitem
\bibitem[Yang \latin{et~al.}(2015)Yang, Ding, and Urban]{yang15}
Yang,~Y.; Ding,~X.; Urban,~M.~W. {Chemical and physical aspects of self-healing
  materials}. \emph{Prog. Polym. Sci.} \textbf{2015}, \emph{49-50}, 34 --
  59\relax
\mciteBstWouldAddEndPuncttrue
\mciteSetBstMidEndSepPunct{\mcitedefaultmidpunct}
{\mcitedefaultendpunct}{\mcitedefaultseppunct}\relax
\EndOfBibitem
\bibitem[Li \latin{et~al.}(2022)Li, Cao, Saito, and Sokolov]{Li22}
Li,~B.; Cao,~P.-F.; Saito,~T.; Sokolov,~A.~P. {Intrinsically Self-Healing
  Polymers: From Mechanistic Insight to Current Challenges}. \emph{Chem. Rev.}
  \textbf{2022}, \emph{123}, 701--735\relax
\mciteBstWouldAddEndPuncttrue
\mciteSetBstMidEndSepPunct{\mcitedefaultmidpunct}
{\mcitedefaultendpunct}{\mcitedefaultseppunct}\relax
\EndOfBibitem
\bibitem[Dufort and Tibbitt(2019)Dufort, and Tibbitt]{2019.Dufort}
Dufort,~B.~M.; Tibbitt,~M. {Design of moldable hydrogels for biomedical
  applications using dynamic covalent boronic esters}. \emph{Mater. Today
  Chem.} \textbf{2019}, \emph{12}, 16--33\relax
\mciteBstWouldAddEndPuncttrue
\mciteSetBstMidEndSepPunct{\mcitedefaultmidpunct}
{\mcitedefaultendpunct}{\mcitedefaultseppunct}\relax
\EndOfBibitem
\bibitem[Zhang \latin{et~al.}(2014)Zhang, Zhang, Chai, Xue, Hao, and
  Zheng]{zhang14}
Zhang,~L.; Zhang,~L.; Chai,~L.; Xue,~P.; Hao,~W.; Zheng,~H. {A coordinatively
  cross-linked polymeric network as a functional binder for high-performance
  silicon submicro-particle\ anodes in lithium-ion batteries}. \emph{J. Mater.
  Chem. A} \textbf{2014}, \emph{2}, 19036--19045\relax
\mciteBstWouldAddEndPuncttrue
\mciteSetBstMidEndSepPunct{\mcitedefaultmidpunct}
{\mcitedefaultendpunct}{\mcitedefaultseppunct}\relax
\EndOfBibitem
\bibitem[Kwon \latin{et~al.}(2015)Kwon, Jeong, Deniz, AlQaradawi, Choi, and
  Coskun]{Kwon.2015}
Kwon,~T.-w.; Jeong,~Y.~K.; Deniz,~E.; AlQaradawi,~S.~Y.; Choi,~J.~W.;
  Coskun,~A. {Dynamic Cross-Linking of Polymeric Binders Based on Host–Guest
  Interactions for Silicon Anodes in Lithium Ion Batteri\ es}. \emph{ACS Nano}
  \textbf{2015}, \emph{9}, 11317--11324\relax
\mciteBstWouldAddEndPuncttrue
\mciteSetBstMidEndSepPunct{\mcitedefaultmidpunct}
{\mcitedefaultendpunct}{\mcitedefaultseppunct}\relax
\EndOfBibitem
\bibitem[Yue \latin{et~al.}(2016)Yue, Ma, Zhang, Zhao, Dong, Liu, Cui, and
  Chen]{Yue06}
Yue,~L.; Ma,~J.; Zhang,~J.; Zhao,~J.; Dong,~S.; Liu,~Z.; Cui,~G.; Chen,~L. {All
  solid-state polymer electrolytes for high-performance lithium ion batteries}.
  \emph{Energy Storage Mater.} \textbf{2016}, \emph{5}, 139--164\relax
\mciteBstWouldAddEndPuncttrue
\mciteSetBstMidEndSepPunct{\mcitedefaultmidpunct}
{\mcitedefaultendpunct}{\mcitedefaultseppunct}\relax
\EndOfBibitem
\bibitem[Mindemark \latin{et~al.}(2018)Mindemark, Lacey, Bowden, and
  Brandell]{2018_MindemarkLaceyEtAl}
Mindemark,~J.; Lacey,~M.~J.; Bowden,~T.; Brandell,~D. Beyond
  {PEO}{\textemdash}Alternative Host Materials for Li$^+$-Conducting Solid
  Polymer Electrolytes. \emph{Prog. Polym. Sci.} \textbf{2018}, \emph{81},
  114--143\relax
\mciteBstWouldAddEndPuncttrue
\mciteSetBstMidEndSepPunct{\mcitedefaultmidpunct}
{\mcitedefaultendpunct}{\mcitedefaultseppunct}\relax
\EndOfBibitem
\bibitem[Zhang \latin{et~al.}(2023)Zhang, Cheng, Chen, Chan, Cai, Carvalho,
  Marchiori, Brandell, Araujo, Chen, Ji, Feng, Goloviznina, \~Serva, Salanne,
  Mandai, Hosaka, Alhanash, Johansson, \~Qiu, Xiao, Eikerling, Jinnouchi,
  Melander, Kastlunger, Bouzid\, Pasquarello, Shin, Kim, Kim, Schwarz, and
  Sundararaman]{zhang23}
Zhang,~C. \latin{et~al.}  {2023 roadmap on molecular modelling of
  electrochemical energy materials}. \emph{JPhys Energy} \textbf{2023},
  \emph{5}, 041501\relax
\mciteBstWouldAddEndPuncttrue
\mciteSetBstMidEndSepPunct{\mcitedefaultmidpunct}
{\mcitedefaultendpunct}{\mcitedefaultseppunct}\relax
\EndOfBibitem
\bibitem[France-Lanord and Grossman(2019)France-Lanord, and
  Grossman]{2019_FranceLanordGrossman}
France-Lanord,~A.; Grossman,~J.~C. Correlations from Ion Pairing and the
  Nernst-Einstein Equation. \emph{Phys. Rev. Lett.} \textbf{2019}, \emph{122},
  136001\relax
\mciteBstWouldAddEndPuncttrue
\mciteSetBstMidEndSepPunct{\mcitedefaultmidpunct}
{\mcitedefaultendpunct}{\mcitedefaultseppunct}\relax
\EndOfBibitem
\bibitem[Fong \latin{et~al.}(2019)Fong, Self, Diederichsen, Wood, McCloskey,
  and Persson]{fong19}
Fong,~K.~D.; Self,~J.; Diederichsen,~K.~M.; Wood,~B.~M.; McCloskey,~B.~D.;
  Persson,~K. n.~A. {Ion Transport and the True Transference Number in
  Nonaqueous Polyelectrolyte Solutions for Lithium Ion Batteries.} \emph{ACS
  Cent. Sci.} \textbf{2019}, \emph{5}, 1250 -- 1260\relax
\mciteBstWouldAddEndPuncttrue
\mciteSetBstMidEndSepPunct{\mcitedefaultmidpunct}
{\mcitedefaultendpunct}{\mcitedefaultseppunct}\relax
\EndOfBibitem
\bibitem[Zhang \latin{et~al.}(2020)Zhang, Wheatle, Krajniak, Keith, and
  Ganesan]{zhang2020}
Zhang,~Z.; Wheatle,~B.~K.; Krajniak,~J.; Keith,~J.~R.; Ganesan,~V. {Ion
  Mobilities, Transference Numbers, and Inverse Haven Ratios of Polymeric Ionic
  Liquids}. \emph{ACS Macro Lett.} \textbf{2020}, \emph{9}, 84--89\relax
\mciteBstWouldAddEndPuncttrue
\mciteSetBstMidEndSepPunct{\mcitedefaultmidpunct}
{\mcitedefaultendpunct}{\mcitedefaultseppunct}\relax
\EndOfBibitem
\bibitem[Shen and Hall(2020)Shen, and Hall]{Shen.2020}
Shen,~K.-H.; Hall,~L.~M. {Effects of Ion Size and Dielectric Constant on Ion
  Transport and Transference Number in Polymer Electrolytes}.
  \emph{Macromolecules} \textbf{2020}, \emph{53}, 10086--10096\relax
\mciteBstWouldAddEndPuncttrue
\mciteSetBstMidEndSepPunct{\mcitedefaultmidpunct}
{\mcitedefaultendpunct}{\mcitedefaultseppunct}\relax
\EndOfBibitem
\bibitem[Bollinger \latin{et~al.}(2020)Bollinger, Stevens, and
  Frischknecht]{bollinger20}
Bollinger,~J.~A.; Stevens,~M.~J.; Frischknecht,~A.~L. {Quantifying Single-Ion
  Transport in Percolated Ionic Aggregates of Polymer Melts}. \emph{ACS Macro
  Lett.} \textbf{2020}, \emph{9}, 583--587\relax
\mciteBstWouldAddEndPuncttrue
\mciteSetBstMidEndSepPunct{\mcitedefaultmidpunct}
{\mcitedefaultendpunct}{\mcitedefaultseppunct}\relax
\EndOfBibitem
\bibitem[Gudla \latin{et~al.}(2020)Gudla, Zhang, and Brandell]{Gudla2020}
Gudla,~H.; Zhang,~C.; Brandell,~D. {Effects of Solvent Polarity on Li-Ion
  Diffusion in Polymer Electrolytes: An All-atom Molecular Dynamics Study with
  Charge Scaling}. \emph{J. Phys. Chem. B} \textbf{2020}, \emph{124},
  8124--8131\relax
\mciteBstWouldAddEndPuncttrue
\mciteSetBstMidEndSepPunct{\mcitedefaultmidpunct}
{\mcitedefaultendpunct}{\mcitedefaultseppunct}\relax
\EndOfBibitem
\bibitem[Gudla \latin{et~al.}(2021)Gudla, Shao, Phunnarungsi, Brandell, and
  Zhang]{Gudla2021}
Gudla,~H.; Shao,~Y.; Phunnarungsi,~S.; Brandell,~D.; Zhang,~C. {Importance of
  the Ion-Pair Lifetime in Polymer Electrolytes}. \emph{J. Phys. Chem. Lett.}
  \textbf{2021}, \emph{12}, 8460--8464\relax
\mciteBstWouldAddEndPuncttrue
\mciteSetBstMidEndSepPunct{\mcitedefaultmidpunct}
{\mcitedefaultendpunct}{\mcitedefaultseppunct}\relax
\EndOfBibitem
\bibitem[Halat \latin{et~al.}(2022)Halat, Fang, Hickson, Mistry, Reimer,
  Balsara, and Wang]{2022.Halat}
Halat,~D.~M.; Fang,~C.; Hickson,~D.; Mistry,~A.; Reimer,~J.~A.; Balsara,~N.~P.;
  Wang,~R. {Electric-Field-Induced Spatially Dynamic Heterogeneity of Solvent
  Motion and Cation Transference in Electrolytes}. \emph{Phys. Rev. Lett.}
  \textbf{2022}, \emph{128}, 198002\relax
\mciteBstWouldAddEndPuncttrue
\mciteSetBstMidEndSepPunct{\mcitedefaultmidpunct}
{\mcitedefaultendpunct}{\mcitedefaultseppunct}\relax
\EndOfBibitem
\bibitem[Shao \latin{et~al.}(2022)Shao, Gudla, Brandell, and Zhang]{Shao2022}
Shao,~Y.; Gudla,~H.; Brandell,~D.; Zhang,~C. {Transference Number in Polymer
  Electrolytes: Mind the Reference-Frame Gap}. \emph{J. Am. Chem. Soc.}
  \textbf{2022}, \emph{144}, 7583--7587\relax
\mciteBstWouldAddEndPuncttrue
\mciteSetBstMidEndSepPunct{\mcitedefaultmidpunct}
{\mcitedefaultendpunct}{\mcitedefaultseppunct}\relax
\EndOfBibitem
\bibitem[Shao and Zhang(2023)Shao, and Zhang]{Shao2023}
Shao,~Y.; Zhang,~C. {Bruce–Vincent transference numbers from molecular
  dynamics simulations}. \emph{J. Chem. Phys.} \textbf{2023}, \emph{158},
  161104\relax
\mciteBstWouldAddEndPuncttrue
\mciteSetBstMidEndSepPunct{\mcitedefaultmidpunct}
{\mcitedefaultendpunct}{\mcitedefaultseppunct}\relax
\EndOfBibitem
\bibitem[Mogurampelly \latin{et~al.}(2016)Mogurampelly, Sethuraman, Pryamitsyn,
  and Ganesan]{mogurampelly16}
Mogurampelly,~S.; Sethuraman,~V.; Pryamitsyn,~V.; Ganesan,~V. {Influence of
  nanoparticle-ion and nanoparticle-polymer interactions on ion transport and
  viscoelastic properties of p\ olymer electrolytes}. \emph{J. Chem. Phys.}
  \textbf{2016}, \emph{144}, 154905\relax
\mciteBstWouldAddEndPuncttrue
\mciteSetBstMidEndSepPunct{\mcitedefaultmidpunct}
{\mcitedefaultendpunct}{\mcitedefaultseppunct}\relax
\EndOfBibitem
\bibitem[Verners \latin{et~al.}(2016)Verners, Thijsse, Duin, and
  Simone]{verners16}
Verners,~O.; Thijsse,~B.~J.; Duin,~A. C.~v.; Simone,~A. {Salt concentration
  effects on mechanical properties of LiPF6/poly(propylene glycol) diacrylate
  solid electrolyte: Ins\ ights from reactive molecular dynamics simulations}.
  \emph{Electrochim. Acta} \textbf{2016}, \emph{221}, 115--123\relax
\mciteBstWouldAddEndPuncttrue
\mciteSetBstMidEndSepPunct{\mcitedefaultmidpunct}
{\mcitedefaultendpunct}{\mcitedefaultseppunct}\relax
\EndOfBibitem
\bibitem[Sampath and Hall(2017)Sampath, and Hall]{sampath17}
Sampath,~J.; Hall,~L.~M. {Impact of ionic aggregate structure on ionomer
  mechanical properties from coarse-grained molecular dynamics simulatio\ ns}.
  \emph{J. Chem. Phys.} \textbf{2017}, \emph{147}, 134901\relax
\mciteBstWouldAddEndPuncttrue
\mciteSetBstMidEndSepPunct{\mcitedefaultmidpunct}
{\mcitedefaultendpunct}{\mcitedefaultseppunct}\relax
\EndOfBibitem
\bibitem[Verners \latin{et~al.}(2018)Verners, Lyulin, and Simone]{Verners18}
Verners,~O.; Lyulin,~A.~V.; Simone,~A. {Salt concentration dependence of the
  mechanical properties of LiPF6/poly(propylene glycol) acrylate electrolyte at
  a \ graphitic carbon interface: A reactive molecular dynamics study}.
  \emph{J. Polym. Sci., Part B: Polym. Phys.} \textbf{2018}, \emph{56},
  718--730\relax
\mciteBstWouldAddEndPuncttrue
\mciteSetBstMidEndSepPunct{\mcitedefaultmidpunct}
{\mcitedefaultendpunct}{\mcitedefaultseppunct}\relax
\EndOfBibitem
\bibitem[Sampath and Hall(2018)Sampath, and Hall]{sampath18}
Sampath,~J.; Hall,~L.~M. {Impact of ion content and electric field on
  mechanical properties of coarse-grained ionomers}. \emph{J. Chem. Phys.}
  \textbf{2018}, \emph{149}, 163313\relax
\mciteBstWouldAddEndPuncttrue
\mciteSetBstMidEndSepPunct{\mcitedefaultmidpunct}
{\mcitedefaultendpunct}{\mcitedefaultseppunct}\relax
\EndOfBibitem
\bibitem[Brinson and Brinson(2015)Brinson, and Brinson]{Brinson2015}
Brinson,~H.~F.; Brinson,~L.~C. \emph{Polymer Engineering Science and
  Viscoelasticity: An Introduction}; 2015; pp 1--482\relax
\mciteBstWouldAddEndPuncttrue
\mciteSetBstMidEndSepPunct{\mcitedefaultmidpunct}
{\mcitedefaultendpunct}{\mcitedefaultseppunct}\relax
\EndOfBibitem
\bibitem[Suter and Eichinger(2002)Suter, and Eichinger]{Suter2002}
Suter,~U.; Eichinger,~B. {Estimating elastic constants by averaging over
  simulated structures}. \emph{Polymer} \textbf{2002}, \emph{43},
  575--582\relax
\mciteBstWouldAddEndPuncttrue
\mciteSetBstMidEndSepPunct{\mcitedefaultmidpunct}
{\mcitedefaultendpunct}{\mcitedefaultseppunct}\relax
\EndOfBibitem
\bibitem[Clancy \latin{et~al.}(2009)Clancy, Frankland, Hinkley, and
  Gates]{Clancy2009}
Clancy,~T.~C.; Frankland,~S.~J.; Hinkley,~J.~A.; Gates,~T.~S. {Molecular
  modeling for calculation of mechanical properties of epoxies with moisture
  ingress}. \emph{Polymer} \textbf{2009}, \emph{50}, 2736--2742\relax
\mciteBstWouldAddEndPuncttrue
\mciteSetBstMidEndSepPunct{\mcitedefaultmidpunct}
{\mcitedefaultendpunct}{\mcitedefaultseppunct}\relax
\EndOfBibitem
\bibitem[David \latin{et~al.}(2019)David, {De Nicola}, Tartaglino, Milano, and
  Raos]{David2019}
David,~A.; {De Nicola},~A.; Tartaglino,~U.; Milano,~G.; Raos,~G.
  {Viscoelasticity of Short Polymer Liquids from Atomistic Simulations}.
  \emph{J. Electrochem. Soc.} \textbf{2019}, \emph{166}, B3246--B3256\relax
\mciteBstWouldAddEndPuncttrue
\mciteSetBstMidEndSepPunct{\mcitedefaultmidpunct}
{\mcitedefaultendpunct}{\mcitedefaultseppunct}\relax
\EndOfBibitem
\bibitem[Liu \latin{et~al.}(2012)Liu, Maginn, Visser, Bridges, and
  Fox]{2012.Liu20a}
Liu,~H.; Maginn,~E.; Visser,~A.~E.; Bridges,~N.~J.; Fox,~E.~B. {Thermal and
  Transport Properties of Six Ionic Liquids: An Experimental and Molecular
  Dynamics Study}. \emph{Ind. Eng. Chem. Res.} \textbf{2012}, \emph{51},
  7242--7254\relax
\mciteBstWouldAddEndPuncttrue
\mciteSetBstMidEndSepPunct{\mcitedefaultmidpunct}
{\mcitedefaultendpunct}{\mcitedefaultseppunct}\relax
\EndOfBibitem
\bibitem[Ram{\'{i}}rez \latin{et~al.}(2010)Ram{\'{i}}rez, Sukumaran,
  Vorselaars, and Likhtman]{Ramirez2010}
Ram{\'{i}}rez,~J.; Sukumaran,~S.~K.; Vorselaars,~B.; Likhtman,~A.~E. {Efficient
  on the fly calculation of time correlation functions in computer
  simulations}. \emph{J. Chem. Phys.} \textbf{2010}, \emph{133}\relax
\mciteBstWouldAddEndPuncttrue
\mciteSetBstMidEndSepPunct{\mcitedefaultmidpunct}
{\mcitedefaultendpunct}{\mcitedefaultseppunct}\relax
\EndOfBibitem
\bibitem[{Paul Müller}(2012)]{multau}
{Paul Müller}, Python multiple-tau algorithm (Version 0.3.3).
  \url{https://pypi.python.org/pypi/multipletau/}, 2012; Accessed:
  2023-09-01\relax
\mciteBstWouldAddEndPuncttrue
\mciteSetBstMidEndSepPunct{\mcitedefaultmidpunct}
{\mcitedefaultendpunct}{\mcitedefaultseppunct}\relax
\EndOfBibitem
\bibitem[Adeyemi(2022)]{Adeyemi2022}
Adeyemi,~O. {Molecular dynamics investigation of viscoelastic properties of
  polymer melts}. Ph.D.\ thesis, 2022\relax
\mciteBstWouldAddEndPuncttrue
\mciteSetBstMidEndSepPunct{\mcitedefaultmidpunct}
{\mcitedefaultendpunct}{\mcitedefaultseppunct}\relax
\EndOfBibitem
\bibitem[Adeyemi \latin{et~al.}(2022)Adeyemi, Zhu, and Xi]{Adeyemi2022a}
Adeyemi,~O.; Zhu,~S.; Xi,~L. {Equilibrium and non-equilibrium molecular
  dynamics approaches for the linear viscoelasticity of polymer melts}.
  \emph{Phys. Fluids} \textbf{2022}, \emph{34}\relax
\mciteBstWouldAddEndPuncttrue
\mciteSetBstMidEndSepPunct{\mcitedefaultmidpunct}
{\mcitedefaultendpunct}{\mcitedefaultseppunct}\relax
\EndOfBibitem
\bibitem[Wang \latin{et~al.}(2004)Wang, Wolf, Caldwell, Kollman, and
  Case]{Wang2004}
Wang,~J.; Wolf,~R.~M.; Caldwell,~J.~W.; Kollman,~P.~A.; Case,~D.~A.
  {Development and testing of a general amber force field}. \emph{J. Comput.
  Chem.} \textbf{2004}, \emph{25}, 1157--1174\relax
\mciteBstWouldAddEndPuncttrue
\mciteSetBstMidEndSepPunct{\mcitedefaultmidpunct}
{\mcitedefaultendpunct}{\mcitedefaultseppunct}\relax
\EndOfBibitem
\bibitem[Thompson \latin{et~al.}(2022)Thompson, Aktulga, Berger, Bolintineanu,
  Brown, Crozier, in~'t Veld, Kohlmeyer, Moore, Nguyen, Shan, Stevens,
  Tranchida, Trott, and Plimpton]{LAMMPS}
Thompson,~A.~P.; Aktulga,~H.~M.; Berger,~R.; Bolintineanu,~D.~S.; Brown,~W.~M.;
  Crozier,~P.~S.; in~'t Veld,~P.~J.; Kohlmeyer,~A.; Moore,~S.~G.;
  Nguyen,~T.~D.; Shan,~R.; Stevens,~M.~J.; Tranchida,~J.; Trott,~C.;
  Plimpton,~S.~J. {LAMMPS} - a flexible simulation tool for particle-based
  materials modeling at the atomic, meso, and continuum scales. \emph{Comp.
  Phys. Comm.} \textbf{2022}, \emph{271}, 108171\relax
\mciteBstWouldAddEndPuncttrue
\mciteSetBstMidEndSepPunct{\mcitedefaultmidpunct}
{\mcitedefaultendpunct}{\mcitedefaultseppunct}\relax
\EndOfBibitem
\bibitem[Abraham \latin{et~al.}(2015)Abraham, Murtola, Schulz, P{\'{a}}ll,
  Smith, Hess, and Lindah]{Abraham2015}
Abraham,~M.~J.; Murtola,~T.; Schulz,~R.; P{\'{a}}ll,~S.; Smith,~J.~C.;
  Hess,~B.; Lindah,~E. {GROMACS: High performance molecular simulations through
  multi-level parallelism from laptops to supercomputers}. \emph{SoftwareX}
  \textbf{2015}, \emph{1-2}, 19--25\relax
\mciteBstWouldAddEndPuncttrue
\mciteSetBstMidEndSepPunct{\mcitedefaultmidpunct}
{\mcitedefaultendpunct}{\mcitedefaultseppunct}\relax
\EndOfBibitem
\bibitem[Evans and Holian(1985)Evans, and Holian]{Evans85}
Evans,~D.~J.; Holian,~B.~L. {The Nose–Hoover thermostat}. \emph{J. Chem.
  Phys.} \textbf{1985}, \emph{83}, 4069--4074\relax
\mciteBstWouldAddEndPuncttrue
\mciteSetBstMidEndSepPunct{\mcitedefaultmidpunct}
{\mcitedefaultendpunct}{\mcitedefaultseppunct}\relax
\EndOfBibitem
\bibitem[Evans and Morriss(1984)Evans, and Morriss]{Evans84}
Evans,~D.~J.; Morriss,~G.~P. Nonlinear-response theory for steady planar
  Couette flow. \emph{Phys. Rev. A} \textbf{1984}, \emph{30}, 1528--1530\relax
\mciteBstWouldAddEndPuncttrue
\mciteSetBstMidEndSepPunct{\mcitedefaultmidpunct}
{\mcitedefaultendpunct}{\mcitedefaultseppunct}\relax
\EndOfBibitem
\bibitem[Nosé and Klein(1983)Nosé, and Klein]{Noose83}
Nosé,~S.; Klein,~M. Constant pressure molecular dynamics for molecular
  systems. \emph{Mol. Phys.} \textbf{1983}, \emph{50}, 1055--1076\relax
\mciteBstWouldAddEndPuncttrue
\mciteSetBstMidEndSepPunct{\mcitedefaultmidpunct}
{\mcitedefaultendpunct}{\mcitedefaultseppunct}\relax
\EndOfBibitem
\bibitem[Wu(2019)]{Wu2019}
Wu,~C. {Bulk modulus of poly(ethylene oxide) simulated using the systematically
  coarse-grained model}. \emph{Comput. Mater. Sci.} \textbf{2019}, \emph{156},
  89--95\relax
\mciteBstWouldAddEndPuncttrue
\mciteSetBstMidEndSepPunct{\mcitedefaultmidpunct}
{\mcitedefaultendpunct}{\mcitedefaultseppunct}\relax
\EndOfBibitem
\bibitem[Greaves \latin{et~al.}(2011)Greaves, Greer, Lakes, and
  Rouxel]{Greaves2011}
Greaves,~G.~N.; Greer,~A.~L.; Lakes,~R.~S.; Rouxel,~T. {Poisson's ratio and
  modern materials}. \emph{Nat. Mater.} \textbf{2011}, \emph{10},
  823--837\relax
\mciteBstWouldAddEndPuncttrue
\mciteSetBstMidEndSepPunct{\mcitedefaultmidpunct}
{\mcitedefaultendpunct}{\mcitedefaultseppunct}\relax
\EndOfBibitem
\bibitem[Li \latin{et~al.}(2020)Li, Zhu, Yao, Qian, Zhang, Yan, and
  Wang]{Li2020}
Li,~J.; Zhu,~K.; Yao,~Z.; Qian,~G.; Zhang,~J.; Yan,~K.; Wang,~J. {A promising
  composite solid electrolyte incorporating LLZO into PEO/PVDF matrix for
  all-solid-state lithium-ion batteries}. \emph{Ionics} \textbf{2020},
  \emph{26}, 1101--1108\relax
\mciteBstWouldAddEndPuncttrue
\mciteSetBstMidEndSepPunct{\mcitedefaultmidpunct}
{\mcitedefaultendpunct}{\mcitedefaultseppunct}\relax
\EndOfBibitem
\bibitem[Jee \latin{et~al.}(2013)Jee, Lee, Lee, and Lee]{Jee2013}
Jee,~A.-Y.; Lee,~H.; Lee,~Y.; Lee,~M. {Determination of the elastic modulus of
  poly(ethylene oxide) using a photoisomerizing dye}. \emph{Chem. Phys.}
  \textbf{2013}, \emph{422}, 246--250\relax
\mciteBstWouldAddEndPuncttrue
\mciteSetBstMidEndSepPunct{\mcitedefaultmidpunct}
{\mcitedefaultendpunct}{\mcitedefaultseppunct}\relax
\EndOfBibitem
\bibitem[Angulakshmi \latin{et~al.}(2014)Angulakshmi, Kumar, Kulandainathan,
  and Stephan]{Angulakshmi2014}
Angulakshmi,~N.; Kumar,~R.~S.; Kulandainathan,~M.~A.; Stephan,~A.~M. {Composite
  Polymer Electrolytes Encompassing Metal Organic Frame Works: A New Strategy
  for All-Solid-State Lithium Batteries}. \emph{J. Phys. Chem. C}
  \textbf{2014}, \emph{118}, 24240--24247\relax
\mciteBstWouldAddEndPuncttrue
\mciteSetBstMidEndSepPunct{\mcitedefaultmidpunct}
{\mcitedefaultendpunct}{\mcitedefaultseppunct}\relax
\EndOfBibitem
\bibitem[Karuppasamy \latin{et~al.}(2016)Karuppasamy, Rhee, Reddy, Gupta, Mitu,
  Polu, and {Sahaya Shajan}]{Karuppasamy2016}
Karuppasamy,~K.; Rhee,~H.~W.; Reddy,~P.~A.; Gupta,~D.; Mitu,~L.; Polu,~A.~R.;
  {Sahaya Shajan},~X. {Ionic liquid incorporated nanocomposite polymer
  electrolytes for rechargeable lithium ion battery: A way to achieve improved
  electrochemical and interfacial properties}. \emph{J. Ind. Eng. Chem.}
  \textbf{2016}, \emph{40}, 168--176\relax
\mciteBstWouldAddEndPuncttrue
\mciteSetBstMidEndSepPunct{\mcitedefaultmidpunct}
{\mcitedefaultendpunct}{\mcitedefaultseppunct}\relax
\EndOfBibitem
\bibitem[Karuppasamy \latin{et~al.}(2017)Karuppasamy, Kim, Kang, Prasanna, and
  Rhee]{Karuppasamy2017}
Karuppasamy,~K.; Kim,~D.; Kang,~Y.~H.; Prasanna,~K.; Rhee,~H.~W. {Improved
  electrochemical, mechanical and transport properties of novel lithium
  bisnonafluoro-1-butanesulfonimidate (LiBNFSI) based solid polymer
  electrolytes for rechargeable lithium ion batteries}. \emph{J. Ind. Eng.
  Chem.} \textbf{2017}, \emph{52}, 224--234\relax
\mciteBstWouldAddEndPuncttrue
\mciteSetBstMidEndSepPunct{\mcitedefaultmidpunct}
{\mcitedefaultendpunct}{\mcitedefaultseppunct}\relax
\EndOfBibitem
\bibitem[Lee \latin{et~al.}(2019)Lee, Howell, Rottmayer, Boeckl, and
  Huang]{Lee2019}
Lee,~J.; Howell,~T.; Rottmayer,~M.; Boeckl,~J.; Huang,~H. {Free-Standing
  PEO/LiTFSI/LAGP Composite Electrolyte Membranes for Applications to Flexible
  Solid-State Lithium-Based Batteries}. \emph{J. Electrochem. Soc.}
  \textbf{2019}, \emph{166}, A416--A422\relax
\mciteBstWouldAddEndPuncttrue
\mciteSetBstMidEndSepPunct{\mcitedefaultmidpunct}
{\mcitedefaultendpunct}{\mcitedefaultseppunct}\relax
\EndOfBibitem
\bibitem[Ye \latin{et~al.}(2015)Ye, Wang, Bi, Xue, Xue, Zhou, Xie, and
  Mai]{Ye2015}
Ye,~Y.-S.; Wang,~H.; Bi,~S.-G.; Xue,~Y.; Xue,~Z.-G.; Zhou,~X.-P.; Xie,~X.-L.;
  Mai,~Y.-W. {High performance composite polymer electrolytes using polymeric
  ionic liquid-functionalized graphene molecular brushes}. \emph{J. Mater.
  Chem. A} \textbf{2015}, \emph{3}, 18064--18073\relax
\mciteBstWouldAddEndPuncttrue
\mciteSetBstMidEndSepPunct{\mcitedefaultmidpunct}
{\mcitedefaultendpunct}{\mcitedefaultseppunct}\relax
\EndOfBibitem
\bibitem[Likhtman \latin{et~al.}(2007)Likhtman, Sukumaran, and
  Ramirez]{Likhtman2007}
Likhtman,~A.~E.; Sukumaran,~S.~K.; Ramirez,~J. {Linear viscoelasticity from
  molecular dynamics simulation of entangled polymers}. \emph{Macromolecules}
  \textbf{2007}, \emph{40}, 6748--6757\relax
\mciteBstWouldAddEndPuncttrue
\mciteSetBstMidEndSepPunct{\mcitedefaultmidpunct}
{\mcitedefaultendpunct}{\mcitedefaultseppunct}\relax
\EndOfBibitem
\bibitem[Bakar \latin{et~al.}(2023)Bakar, Darvishi, Aydemir, Yahsi, Tav,
  Menceloglu, and Senses]{Bakar2023}
Bakar,~R.; Darvishi,~S.; Aydemir,~U.; Yahsi,~U.; Tav,~C.; Menceloglu,~Y.~Z.;
  Senses,~E. {Decoding Polymer Architecture Effect on Ion Clustering, Chain
  Dynamics, and Ionic Conductivity in Polymer Electrolytes}. \emph{ACS Appl.
  Energy Mater.} \textbf{2023}, \emph{6}, 4053--4064\relax
\mciteBstWouldAddEndPuncttrue
\mciteSetBstMidEndSepPunct{\mcitedefaultmidpunct}
{\mcitedefaultendpunct}{\mcitedefaultseppunct}\relax
\EndOfBibitem
\bibitem[Niedzwiedz \latin{et~al.}(2008)Niedzwiedz, Wischnewski,
  Pyckhout-Hintzen, Allgaier, Richter, and Faraone]{Niedzwiedz}
Niedzwiedz,~K.; Wischnewski,~A.; Pyckhout-Hintzen,~W.; Allgaier,~J.;
  Richter,~D.; Faraone,~A. {Chain dynamics and viscoelastic properties of
  poly(ethylene oxide)}. \emph{Macromolecules} \textbf{2008}, \emph{41},
  4866--4872\relax
\mciteBstWouldAddEndPuncttrue
\mciteSetBstMidEndSepPunct{\mcitedefaultmidpunct}
{\mcitedefaultendpunct}{\mcitedefaultseppunct}\relax
\EndOfBibitem
\bibitem[Verdier(2003)]{Verdier2003}
Verdier,~C. {Rheological Properties of Living Materials. From Cells to
  Tissues}. \emph{Theor. Med.} \textbf{2003}, \emph{5}, 67--91\relax
\mciteBstWouldAddEndPuncttrue
\mciteSetBstMidEndSepPunct{\mcitedefaultmidpunct}
{\mcitedefaultendpunct}{\mcitedefaultseppunct}\relax
\EndOfBibitem
\bibitem[Behbahani \latin{et~al.}(2021)Behbahani, Schneider, Rissanou,
  Chazirakis, Bačová, Ja\~na, Li, Doxastakis, Polińska, Burkhart, Müller,
  and Harmanda\~ris]{behbahani21}
Behbahani,~A.~F.; Schneider,~L.; Rissanou,~A.; Chazirakis,~A.; Bačová,~P.;
  Ja\~na,~P.~K.; Li,~W.; Doxastakis,~M.; Polińska,~P.; Burkhart,~C.;
  Müller,~M.; Harmanda\~ris,~V.~A. {Dynamics and Rheology of Polymer Melts via
  Hierarchical Atomistic, Coarse-Grained, and Slip-Spring Simulations}.
  \emph{Macromolecules} \textbf{2021}, \emph{54}, 2740--2762\relax
\mciteBstWouldAddEndPuncttrue
\mciteSetBstMidEndSepPunct{\mcitedefaultmidpunct}
{\mcitedefaultendpunct}{\mcitedefaultseppunct}\relax
\EndOfBibitem
\bibitem[Rubinstein and Colby(2016)Rubinstein, and Colby]{Rubinstein}
Rubinstein,~M.; Colby,~R.~H. \emph{{Polymer physics}}, 1st ed.; Oxford
  University Press Oxford: Oxford, 2016\relax
\mciteBstWouldAddEndPuncttrue
\mciteSetBstMidEndSepPunct{\mcitedefaultmidpunct}
{\mcitedefaultendpunct}{\mcitedefaultseppunct}\relax
\EndOfBibitem
\bibitem[Angell \latin{et~al.}(1998)Angell, Imrie, and Ingram]{Angell:1998vw}
Angell,~C.~A.; Imrie,~C.~T.; Ingram,~M.~D. {From simple electrolyte solutions
  through polymer electrolytes to superionic rubbers: some fundamental cons\
  iderations}. \emph{Polym. Int.} \textbf{1998}, \emph{47}, 9--15\relax
\mciteBstWouldAddEndPuncttrue
\mciteSetBstMidEndSepPunct{\mcitedefaultmidpunct}
{\mcitedefaultendpunct}{\mcitedefaultseppunct}\relax
\EndOfBibitem
\end{mcitethebibliography}

\providecommand{\latin}[1]{#1}
\makeatletter
\providecommand{\doi}
  {\begingroup\let\do\@makeother\dospecials
  \catcode`\{=1 \catcode`\}=2 \doi@aux}
\providecommand{\doi@aux}[1]{\endgroup\texttt{#1}}
\makeatother
\providecommand*\mcitethebibliography{\thebibliography}
\csname @ifundefined\endcsname{endmcitethebibliography}
  {\let\endmcitethebibliography\endthebibliography}{}

\end{document}

\newpage
\section{Comparison between LAMMPS and GROMACS simulations with the same force field implementation}

The procedure of computing $T_\textrm{g}$ and the Nernst-Einstein conductivity $\sigma_\textrm{NE}$ can be found in previous publications.\cite{Gudla2020,Shao2022} % XX Cite reference XX. 

\begin{figure}[ht]
    \centering
    \includegraphics[width=0.9\textwidth]{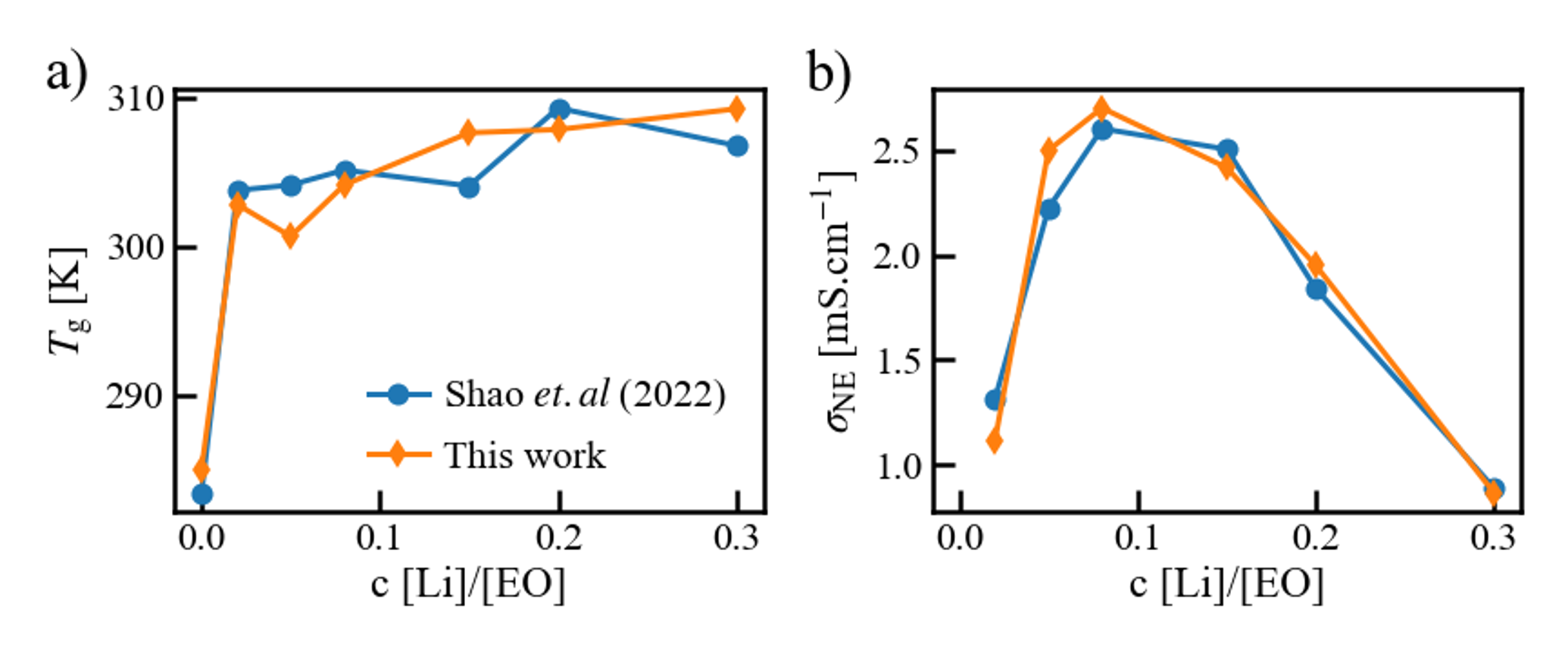}
    \caption{Comparing a) the glass transition temperature $T_\textrm{g}$ and b) the Nernst-Einstein ionic conductivity $\sigma_\textrm{NE}$ from GROMACS and LAMMPS simulations with the same GAFF parameterization.}
    \label{Sfig:gmx-lmp}
\end{figure}

\section{Calculation of end-to-end relaxation time}

\begin{figure}[ht]
    %\centering
    \includegraphics[width=0.9\textwidth]{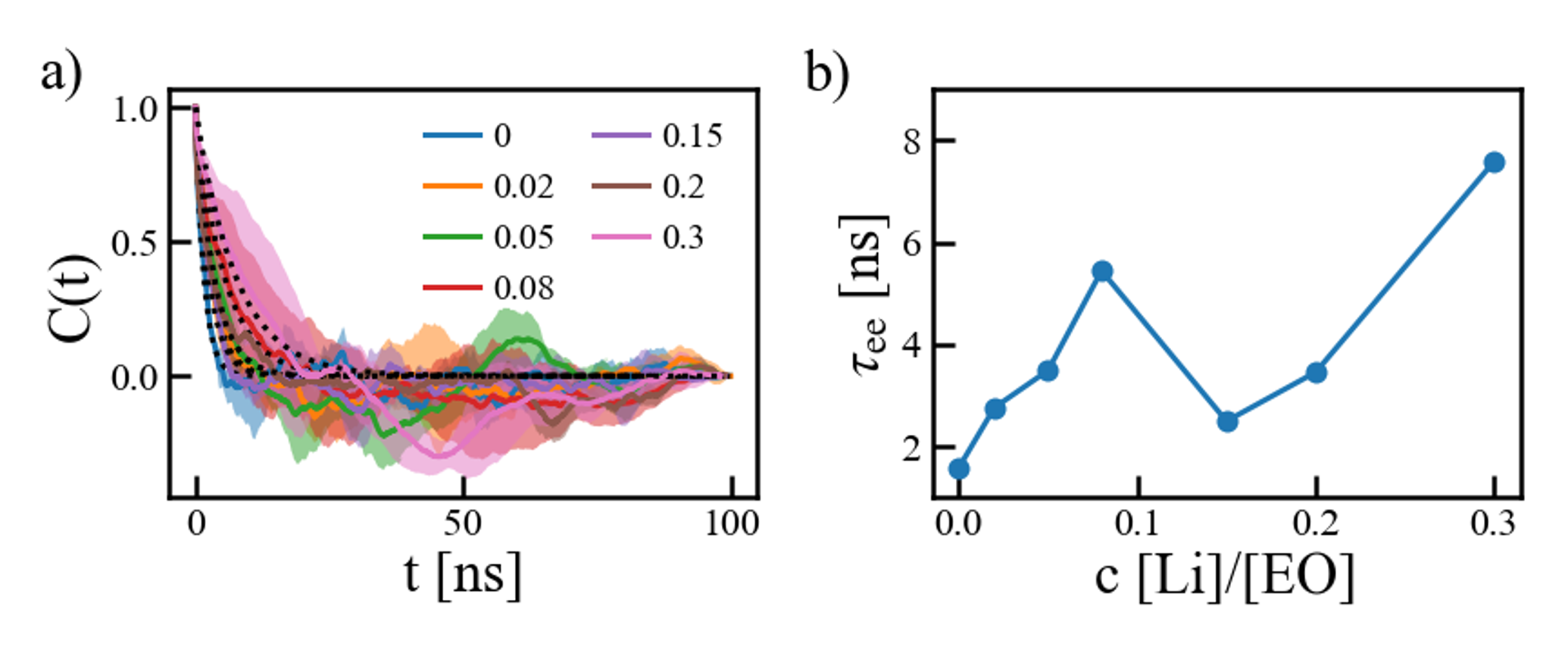}
    \caption{a) Autocorrelation functions of end-to-end distance vector for PEO and PEO-LiTFSI systems at different salt concentrations. Dashed lines are fits to the exponential decay functions. b) End-to-end relaxation time $\tau_{ee}$ as a function of salt concentration.}
    \label{Sfig:ree_tauee}
\end{figure}

The end-to-end distance autocorrelation function $C(t)$ of polymer chains for all systems were plotted in Fig. \ref{Sfig:ree_tauee}, the shaded region corresponds to standard deviation from three different time origins. The relaxation time $\tau_{ee}$ is obtained by fitting $C(t)$ to a simple exponential decay function according to Eq.\ref{Seq:ree_tauee}.
\begin{equation}
    C(t) = \frac{\langle R_{ee}(t_0) \cdot R_{ee}(t_0+t) \rangle }{R_{ee}^2(t_0)} = \exp\left(\frac{-t}{\tau_{ee}}\right)
    \label{Seq:ree_tauee}
\end{equation}

\section{Strain-rate dependence of Lam\'{e}'s constants $\mu$ and $\lambda$}

\begin{figure}[ht]
    \centering
    \includegraphics[width=0.8\textwidth]{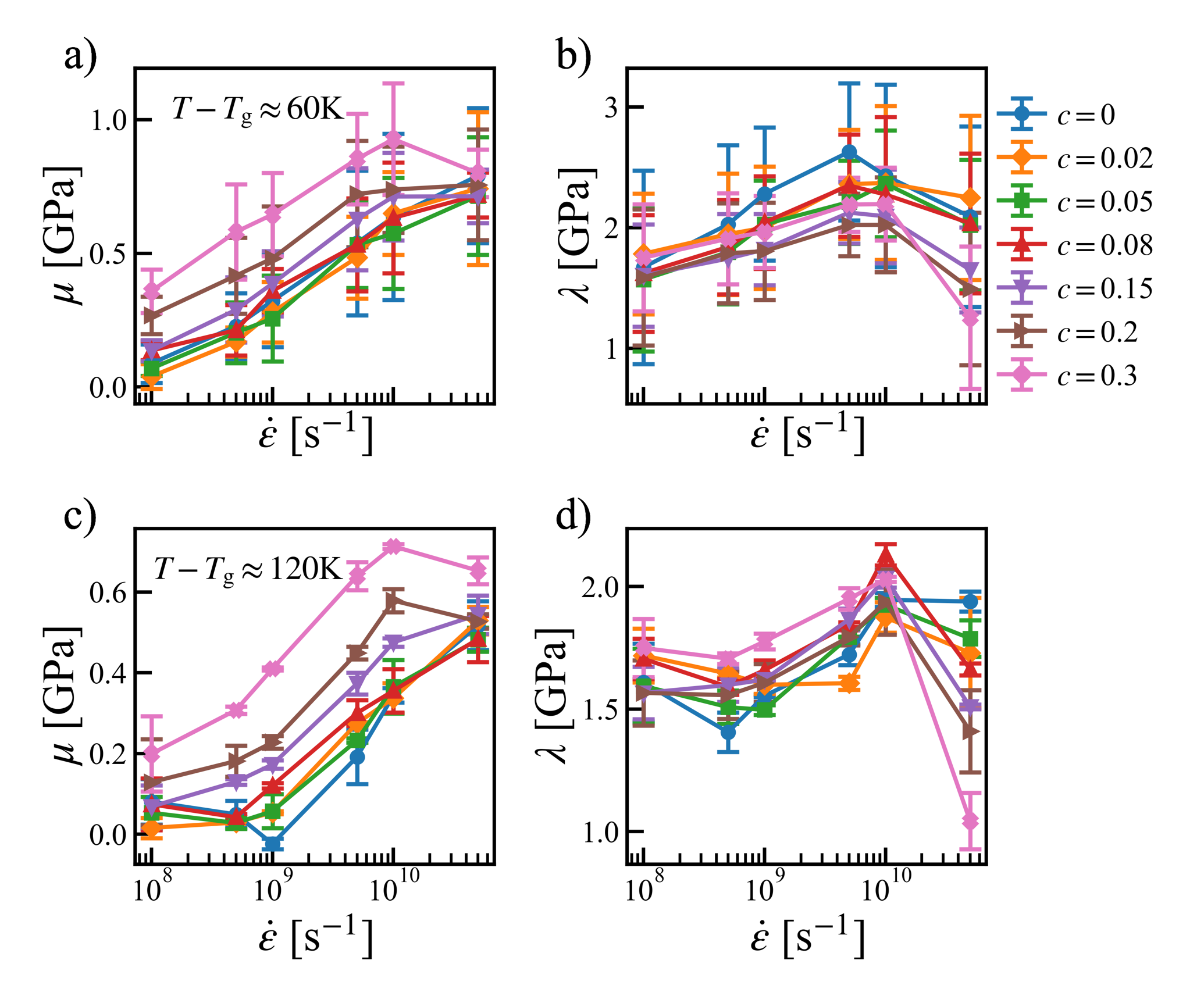}
    \caption{Lam\'{e}'s constants, $\mu$ (a,c) and $\lambda$ (b,d) as function of strain rates at different concentrations and temperatures $T-T_\textrm{g} \approx 60$K and $T-T_\textrm{g} \approx 120$K.}
    \label{Sfig:mu_lamb}
\end{figure}

The Lam\'{e}'s constants, $\mu$ and $\lambda$ calculated for all the systems according to the Eq. 2 in main text at different temperatures, were plotted in Fig. \ref{Sfig:mu_lamb}. Since, Young's modulus $E$ and shear modulus $G$ are directly proportional to $\mu$, the similar strain rate dependence can also been observed there. From Figs. \ref{Sfig:mu_lamb}b,d, the $\lambda$ values are independent of the strain rates, which was also reflected from bulk modulus $B$.  Finally, the Poisson's ratio $\nu$ was inversely dependent on $\mu$, therefore, an opposite strain-rate dependence was observed.   
\newpage
\section{Summary of experimental references on the Youngs' modulus and viscosity}

To compare the Young's modulus and equilibrium viscosities obtained from our simulations with the experimental data, a number of references were found from the literature for the PEO-based electrolytes systems and their details can be found in Table \ref{tab:exp_data}. In Refs.\citenum{Li2020,Angulakshmi2014,Karuppasamy2016,Ye2015}, the Young's modulus was estimated from the stress-strain curves in the elastic regime, i.e. 2-5\% of strain. 
\begin{table}[ht]
    \centering
    \resizebox{\textwidth}{!}
    {\begin{tabular}{ccccccccc}
        \hline\\[1pt]
         Ref. & System & M$_w$  & $c$ & $\Dot{\varepsilon}$  & $T$ & $E$ & $\eta$ \\[5pt]
        \hline\\[1pt]
         Li $et\ al.$\cite{Li2020}& PEO & 300 & - & 0.05 & RT & 160 & -\\
         Jee $et\ al.$\cite{Jee2013}& PEO & 1.1 & - & 0.01 & RT & 31 & -\\
         Angulakshmi $et\ al.$\cite{Angulakshmi2014}
         & PEO:LiTFSI & 300 & 0.05 & 0.00055 & RT & 40 & - & \\
         Karuppasamy $et\ al.$\cite{Karuppasamy2016}& PEO:LiTFSI & 200 & 0.125 & 0.002 & RT & 480 & - & \\
         Karuppasamy $et\ al.$\cite{Karuppasamy2017}& PEO:LiBNFSI & 4000 & 0.05 & 0.01 & RT & 500 & - & \\
         &  &  & 0.0625 &  &  & 27.5 &  & \\
         &  &  & 0.071 &  &  & 12 &  & \\
         Lee $et\ al.$\cite{Lee2019}& PEO:LiTFSI & 400 & 0 & 0.017 & RT & 332.5 & -\\
         &  &  & 0.05 &  &  & 23.2 &  & \\
         Ye $et\ al.$\cite{Ye2015}& PEO:LiCLO$_4$ & 600 & 0.0625 & 0.0014 & RT & 5 & -\\
         Bakar $et al.$\cite{Bakar2023}& PEO:LiTFSI (T$_g\approx$230)& 20 & 0 & - & 348 & - & 36875\\
         &  &  & 0.025 &  &  &  & 20129 & \\
         &  &  & 0.05 &  &  &  & 16150 & \\
         &  &  & 0.085 &  &  &  & 12241 & \\
         &  &  & 0.1 &  &  &  & 8682 & \\
         &  &  & 0.2 &  &  &  & 5075 & \\
        Niedzwiedz $et\ al.$\cite{Niedzwiedz}& PEO (T$_g=$190) & 0.89 & - & - & 308 & - & 0.126\\
        &  &  &  &  & 313 &  & 0.110 & \\
        &  &  &  &  & 318 &  & 0.087 & \\
        &  &  &  &  & 323 &  & 0.071 & \\ 
        \hline\\[1pt]
    \end{tabular}}
    
    \caption{Experimental data extracted from literature. The name of the reference (Ref.), the polymer electrolyte system (System) and glass transition temperature $T_{\textrm{g}}$,K in parentheses, polymer molecular weight (M$_w$) in kg.mol$^{-1}$, salt concentration $c$ [Li]/[EO], strain rate $\Dot{\varepsilon}$ in s$^{-1}$, measured temperature $T$ in K or RT for room temperature, Young's modulus $E$ in MPa and measured equilibrium viscosity $\eta$ in Pa$\cdot$s. }
    \label{tab:exp_data}
\end{table}
\newpage
\section{Calculations of elastic restoration time $\tau_\textrm{res}$}
\addcontentsline{toc}{section}{Calculation of restoration time $\tau_{res}$}

\begin{figure}[ht]
    %\centering
    \includegraphics[width=0.8\textwidth]{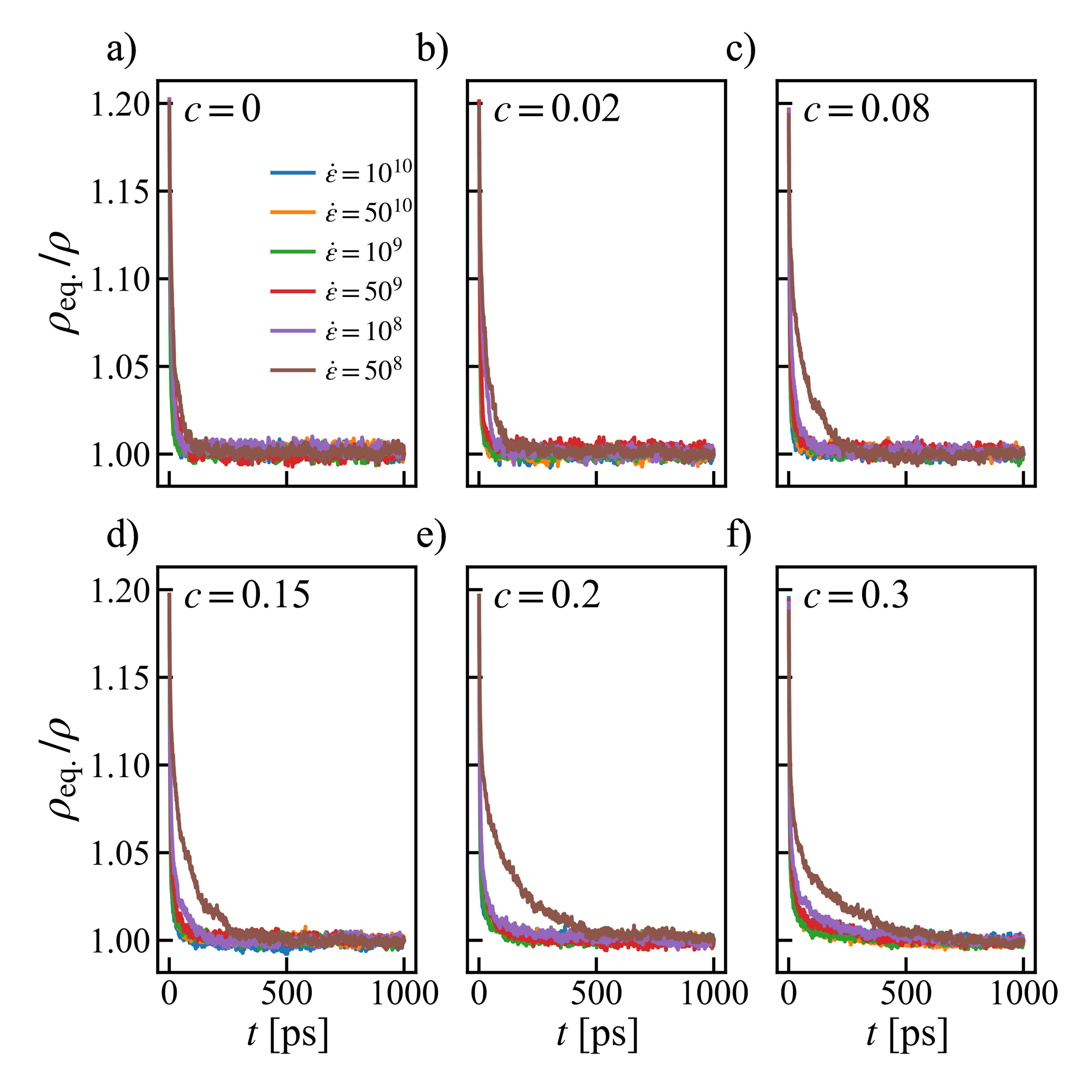}
    \caption{The ratio of equilibrium density $\rho_{\textrm{eq.}}$ with instantaneous density $\rho$ at different strain rates for the neat PEO (a) and PEO-LiTFSI systems at different salt concentrations (b to f).}
    \label{Sfig:rhoeq_rho_tres}
\end{figure}

The final configurations from the uniaxial tensile deformed system (20\% deformation) in all three directions and six strain rates were used in the NPT simulations with Nos\'{e}-Hoover thermostat\cite{Evans85} and barostat\cite{Noose83} for 5 ns and densities were saved every 0.05 ps. The choice of simulation time is sufficient as the densities reach the equilibrium values within this time scale. The ratio of equilibrium density ($\rho_{\textrm{eq.}}$) with instantaneous density ($\rho$) at different strain rates and salt concentrations were plotted in Fig. \ref{Sfig:rhoeq_rho_tres} and were then fitted with a simple exponential decay function to calculate elastic restoration time $\tau_{\textrm{res}}$.

\newpage
% \bibliography{References_HG}

\providecommand{\latin}[1]{#1}
\makeatletter
\providecommand{\doi}
  {\begingroup\let\do\@makeother\dospecials
  \catcode`\{=1 \catcode`\}=2 \doi@aux}
\providecommand{\doi@aux}[1]{\endgroup\texttt{#1}}
\makeatother
\providecommand*\mcitethebibliography{\thebibliography}
\csname @ifundefined\endcsname{endmcitethebibliography}
  {\let\endmcitethebibliography\endthebibliography}{}